\date{\today}

\documentclass[11pt,english]{article}
\usepackage[]{graphicx}

\usepackage[retain-explicit-plus]{siunitx}
\DeclareSIUnit\gpu{gpu}

\usepackage{tikz-cd}
\usetikzlibrary{positioning,fit,calc}
\usetikzlibrary{intersections}
\usetikzlibrary{shapes,arrows}

\usepackage{graphicx}
\usepackage{amsmath,amsfonts}
\usepackage{bm}
\usepackage{url}
\usepackage{verbatim}
\usepackage{subfig}
\usepackage{color}

\graphicspath{{./fig/}}
\DeclareTextSymbol{\degre}{OT1}{23}

\newcommand{\dd}{\mathrm{d}} 
\newcommand{\Uo}{{\mathcal{U}}}
\newcommand{\M}{\mathcal{M}}
\newcommand{\T}{\mathcal{T}}
\newcommand{\C}{\mathcal{C}}
\newcommand{\R}{\mathcal{R}}
\newcommand{\Obs}{\mathcal{O}}
\newcommand{\w}[1]{\vec{\bm{#1}}}
\newcommand{\vv}[1]{\bm{#1}}

\newcommand{\g}[2]{ { \w{ {#1} } \cdot \w{ {#2} } } }
\newcommand{\D}{{\rm d}}
\newcommand{\vD}{\vv{\D}}
\newcommand{\Wg}{W_{\rm geoid}}
\newcommand{\WIAU}{W_{0}^{\rm (IAU)}}
\newcommand{\WEVRF}{W_{0}^{\rm (EVRF2007)}}
\newcommand{\WIHRS}{W_{0}^{\rm (IHRS)}}
\newcommand{\CIAU}{C^{\rm (IAU)}}

\newcommand{\Wzi}{W_0^{(i)}}

\newcommand{\Ci}{C^{(i)}}
\newcommand{\Clev}{C^{\rm (lev)}}
\newcommand{\CGNSS}{C^{\rm (GNSS/geoid)}}
\newcommand{\Hi}{H^{(i)}}
\newcommand{\HNi}{H^{N(i)}}
\newcommand{\HNevrf}{H^{N\rm (EVRF2007)}}
\newcommand{\Ni}{N^{(i)}}
\newcommand{\zi}{\zeta^{(i)}}
\newcommand{\zGNSS}{\zeta_{\rm GNSS}}
\newcommand{\zEGG}{\zeta_{\rm EGG2015}}
\newcommand{\dz}{\delta\zeta}
\newcommand{\dzb}{\overline{\delta\zeta}}
\newcommand{\zzi}{\zeta_0^{(i)}}

\newcommand{\hadj}{h^{\rm (adj)}}
\newcommand{\mss}{\si{\metre\per\square\second}}
\newcommand{\mmss}{\si{\square\metre\per\square\second}}

\newcommand{\pot}[1]{\SI{#1}{\square\metre\per\square\second}}
\newcommand{\grdist}{\delta{g}}
\newcommand{\Xref}[1]{\overline{#1}}
\newcommand{\Xest}[1]{\widetilde{#1}}

\title{Chronometric geodesy: methods and applications}

\author{P. Delva$^{1}$\footnote{Email: pacome.delva@obspm.fr} , H. Denker$^{2}$ and G. Lion$^{3}$\\
$^{1}$ SYRTE\\ Observatoire de Paris, Universit\'e PSL, CNRS, Sorbonne Universit\'e, LNE\\ 61 avenue de l'Observatoire 75014 Paris\\
$^{2}$ Institut f\"ur Erdmessung\\ Leibniz Universit\"at Hannover
(LUH)\\ Schneiderberg 50, 30167 Hannover, Germany\\
$^{3}$ LASTIG LAREG\\ IGN, ENSG, Univ Paris Diderot, Sorbonne
Paris Cit\'e\\ 35 rue H\'el\`ene Brion, 75013 Paris, France
}

\begin{document}
\maketitle


\section{Introduction}

\begin{figure}[t]
	\centering
	\includegraphics[width=\linewidth]{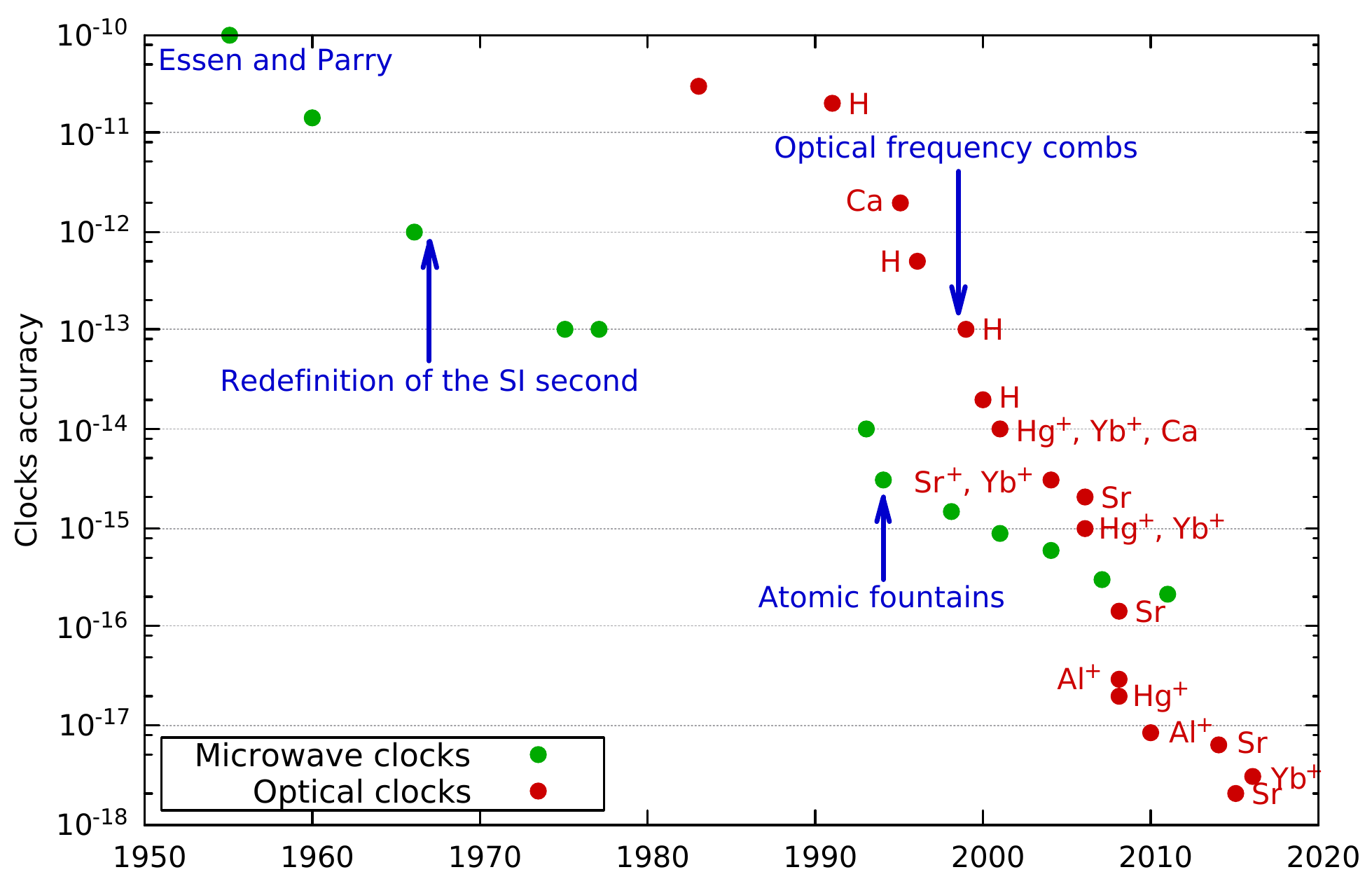}
	\caption{Accuracy records for microwave and optical clocks. From the first Cs clock by Essen and Parry in the 1950's, an order of magnitude was gained every ten years. The advent of optical frequency combs boosted the performances of optical clocks, and they recently surpassed microwave clocks.}
	\label{fig:accuracy}
\end{figure}

The theory of general relativity (GR) was born more than one hundred years ago, and since the beginning has striking prediction success. Einstein proposed three effects for its experimental verification, all verified shortly after their prediction: the perihelion precession of Mercury's orbit, the deflection of light by the Sun, and the gravitational redshift of spectral lines of stars. Other predictions from GR had to wait decades before being confirmed experimentally. It is only in 1959 that the gravitational redshift is confirmed in a lab experiment by Pound, Rebka and Snider~\cite{Pound1959a,Pound1959,Pound1960,Pound1965}. Two gamma-ray emitting iron nuclei at different heights were compared, verifying GR prediction with a relative accuracy of 10\% (and later $<1$\%). In parallel, the era of atomic time began in 1955 with the caesium frequency standard built by Essen and Parry at the National Physical Laboratory (NPL)~\cite{Ramsey2005,Leschiutta2005}. Since then, the accuracy and stability of atomic clocks were constantly ameliorated, with around one order of magnitude gained every ten years (see figure~\ref{fig:accuracy}). 

In this context, the unit of time of the SI, the second, was officially defined with respect to a specified hyperfine transition of the caesium atom in the year 1967--1968\footnote{Resolution 1 of the 13th CGPM~\cite{Terrien1968}.}. Moreover, as local atomic timescales were developed thanks to commercial caesium clocks~\cite{Cutler2005} as well as laboratory caesium standards, there was a need to compare these different timescales in order to build a mean international atomic time, which could be adopted by everyone. This was done at the beginning by USNO and the Bureau International de l'Heure (BIH) with radio time signals, which allowed time comparisons with uncertainties of the order of \SI{1}{ms}~\cite{Guinot2005}. A jump in accuracy occured with the first demonstration from the Hewlett-Packard company of the possibility of using commercial flights to transport its caesium clocks, allowing time transfer with an uncertainty of around \SI{1}{\micro\second}~\cite{Guinot2005}. This eventually led to the famous test of Hafele and Keating~\cite{Hafele1972,Hafele1972b}, who flew four caesium beam clocks around the world on commercial jet flights during October 1971. They predicted and measured the desynchronization of the proper times of these commercial atomic clocks with respect to the USNO atomic scale, and thus verified the gravitational redshift effect with an accuracy of around 12\%\footnote{This is based on the numbers given in table~1 of~\cite{Hafele1972} and table~1 of~\cite{Hafele1972b}: the relative accuracy of the gravitational part of the relativistic shift effect is taken as $(\sqrt{18^2+10^2+7^2}\ {\rm ns}) / (179 \ {\rm ns})$.}.

Following the Hafele and Keating experiment, Briatore and Leschiutta did the first experimental measurements of the gravitational redshift with a direct comparison of ground cesium beam atomic frequency standards~\cite{Briatore1977}. The two clocks were separated in heights by $\Delta h = \SI{3250}{\m}$, predicting a desynchronization of $\Delta t / t \approx g \Delta h / c^2 \approx \SI{30.6}{\ns\per\day}$, where $g$ and $c$ are the local gravity and the velocity of light in vacuum. The measurement gave \SI[separate-uncertainty]{36.5 \pm 5.8}{\ns\per\day}, giving a relative accuracy of around 20\%. Now, we can say that this experiment is amongst the first demonstration of chronometric levelling~(see~section~\ref{sec:chrono}): the clock comparison measured a difference of altitude between the two clocks of \SI[separate-uncertainty]{3880 \pm 620}{\meter}, to be compared with the otherwise measured value of \SI{3250}{\meter}.

Now that the atomic clock accuracy reaches the low $10^{-18}$ in fractional frequency (see figure~\ref{fig:accuracy}), and can be compared to this level over continental distances with optical fibres~(see section~\ref{sec:tech}), the accuracy of chronometric levelling reaches the cm level and begins to be competitive with classical geodetic techniques such as geometric levelling and GNSS/geoid levelling. Moreover, the building of global timescales requires now to take into account these effects to the best possible accuracy. It is the topic of this chapter to explain how atomic clock comparisons and the building of timescales can benefit from the latest developments in physical geodesy for the modelization and realization of the geoid, as well as how classical geodesy could benefit from this new type of observable, which are clock comparisons that are directly linked to gravity potential differences.

In section~\ref{sec:framework} we introduce fundamental concepts of GR concerning the measurement of time, relativistic reference systems and we review the recent literature of chronometric geodesy. In section~\ref{sec:frequ_comp} we introduce the theory of frequency standard comparisons, beginning with the Einstein equivalence principle, followed by the description of the frequency techniques, and finally, we describe clock syntonization and the realization of timescales. Section~\ref{sec:geod} decribes the geodetic methods for determining the gravity potential, namely the geometric levelling approach and the GNSS/geoid approach, as well as considerations about the uncertainties of these methods. In section~\ref{sec:itoc} we describe the European project ITOC where unified relativistic redshift corrections were determined for several clocks in European national metrology institutes. Finally, in section~\ref{sec:high_res} we present numerical simulations exploring what could be the contribution of clock comparisons for the determination of the geoid.

\section{The relativistic framework}
\label{sec:framework}

\subsection{Observers and the space-time manifold}
\label{sec:observers}
The theoretical background of chronometric geodesy is general relativity (GR). In GR theory, space and time are bound together in a continuous entity named \emph{space-time}. Space-time geometry specifies how matter and energy behave, while matter and energy distribution tells how space-time geometry is \emph{curved}. This is a non-linear process and the link between geometry on one side, and matter/energy on the other side is given by the Einstein equations. Gravitation is no longer a force as in Newtonian theory, but the manifestation of the variation of the background geometry. Variations of space-time can be induced by a choice of coordinates, causing inertial effects, which act in a way similar but not equivalent to gravitation. The presence of energy/matter gives rise to curvature of the background geometry, and therefore gravity. However, gravitational effects can never be completely disentangled from inertial effects.

A space-time is formally described by a four-dimensional differentiable manifold $\M$ endowed with a pseudo-Riemannian metric $\bm{g}$. A point of the manifold is called an \emph{event}. Let us define an open subset $\Uo\in\M$ and an event $P$ in this open subset. A \emph{chart} or \emph{coordinate system} $\{x^\alpha\}_{\alpha=0\ldots3}$ can be defined in $\Uo$; it maps point $P\in\Uo$ to a point $\{x^\alpha_P\}\in\mathbb{R}^4$. The four real numbers $x^\alpha_P$ are the \emph{coordinates} of event $P$. Generally a coordinate system cannot be defined on the whole manifold. An atlas is a collection of charts with some properties, which cover the whole manifold. 

As the manifold is smooth, the difference vector $\dd \w{x}$ between two infinitesimally close events may be defined. Then each event $P$ is associated to a vector space $\T_P(\M)$, called the \emph{tangent space}, which contains the set of all possible four-vectors $\dd \w{x}$. A basis $\{\w{e}_\alpha\}$ of the tangent space is usually called a \emph{frame}. The introduced coordinates induce a coordinate basis $\w{e}_\alpha = \{ \w{\partial}_\alpha \}_P \equiv \{ \partial / \partial x^\alpha \}_P$. However, a frame does not need to be associated with any coordinate system. 

The metric tensor $\bm{g}$ is a symmetric bilinear scalar function of two vectors. Given two vectors $\w{v}$ and $\w{w}$, the metric tensor returns a scalar called the dot product: $\bm{g}(\w{v},\w{w}) = \w{v}\cdot\w{w} = \w{w}\cdot\w{v} = \bm{g}(\w{w},\w{v})$. The metric can be characterized by its action on a basis of the tangent space. For example, $g_{\alpha\beta} \equiv \w{\partial}_\alpha\cdot\w{\partial}_\beta$ are the components of the metric tensor in the natural frame associated with the coordinate system $\{x^\alpha\}$. The infinitesimal interval
\begin{equation}
	\dd s^2 \equiv \dd\w{x}\cdot\dd\w{x} = g_{\alpha \beta} \dd x^\alpha \dd x^\beta
	\label{eq:inf_int}
\end{equation}
between two neighboring events is invariant under coordinate transformation. Trajectories of observers are defined by worldlines $\cal{C}$, which are parameterized 1-D curves in the manifold $\M$.

One of the most striking consequences of GR is the fact that coordinates do not have a direct physical interpretation as in the Newtonian theory. Indeed, one has to distinguish proper quantities from coordinate quantities. A proper quantity is the result of a physical measurement in a real or \emph{Gedanken} experiment. It is mathematically described by a scalar, a quantity which is invariant under general coordinate transformations. However, a scalar quantity is not necessarily a physical measurement. The latter needs to be defined with respect to a reference frame adapted to the observer. The proper time $\tau$ of a clock is implicitly defined with the relation
\begin{equation}
\dd s^2 = -c^2 \dd\tau^2 \ ,
\label{eq:dtau}
\end{equation}
where $c$ is a constant velocity characterizing the space-time, which can be identified with the velocity of light in vacuum~\cite{Levy-Leblond1976}. 

Proper time is invariant under general coordinate transformations. The proper times of two different clocks can be compared by means of a time transfer technique, while their proper frequencies can be compared with frequency transfer techniques (see section~\ref{sec:frequ_comp}). Time and frequency comparisons are linked, but both approaches lead to different formalism and experimental techniques. Usually time transfer is more challenging, as it necessitates the knowledge of many instrumental delays with accuracy, while they can often be neglected in the frequency transfer.

When the spatial separation of two clocks is much smaller than the typical length of background curvature of space-time, then curvature effects can safely be neglected. This is a consequence of the Einstein equivalence principle (see e.g.~\cite{Misner1973}). This type of measurement will be termed \emph{local comparison} of clocks. On the contrary, if the distance between both clocks is of the order or bigger than the typical length of the background curvature, the result of the comparison will have curvature perturbations, which depend on both the locations of the clocks and the particular time or frequency transfer technique. This type of measurement will be termed \emph{non-local} or \emph{distant comparison} of clocks.

The relation between proper time and coordinate time can be deduced from equations~(\ref{eq:inf_int})-(\ref{eq:dtau}):
\begin{equation}
	\dd\tau = \frac{1}{c} \sqrt{-\dd\w{x}\cdot\dd\w{x}} \ .
	\label{eq:tau_t1}
\end{equation}
Let us integrate this relation along the worldline ${\cal C}:x^\alpha=f^\alpha(\lambda)$ parameterized by $\lambda$, between two events $A$ and $B$ belonging to $\cal C$ (see figure~\ref{fig:einstein_sync}). The associated tangent vector is $\w{v} = \dot{f}^\alpha\w{\partial}_\alpha$, where $\dot{()}\equiv\dd/\dd\lambda$, and $\dd\w{x}=\w{v}\dd\lambda$, such that:
\begin{equation}
	\tau(A,B) \equiv \int^B_A \dd\tau = \frac{1}{c} \int_{\lambda_A}^{\lambda_B} \dd\lambda\sqrt{-g_{\alpha\beta}\dot{f}^\alpha\dot{f}^\beta} \ ,
	\label{eq:tau_t}
\end{equation}
Parameter $\lambda$ is usually chosen as coordinate $x^0=ct$ in the context of relativistic time and frequency transfer. It is clear from this formula that the proper time elapsed between two events $A$ and $B$ depends on the worldline $\cal C$, i.e. on the trajectory of the observer between these two events. One consequence of GR is that the parameter $\lambda$ cannot be adapted such that both proper and coordinate time be equal everywhere. This is possible only in special relativity where there is no curvature in a well chosen reference system.

\subsection{Simultaneity and synchronization}
\label{sec:sync}

\begin{figure}
   \begin{minipage}[c]{.46\linewidth}
	   \begin{center}
		\includegraphics[width=0.8\linewidth]{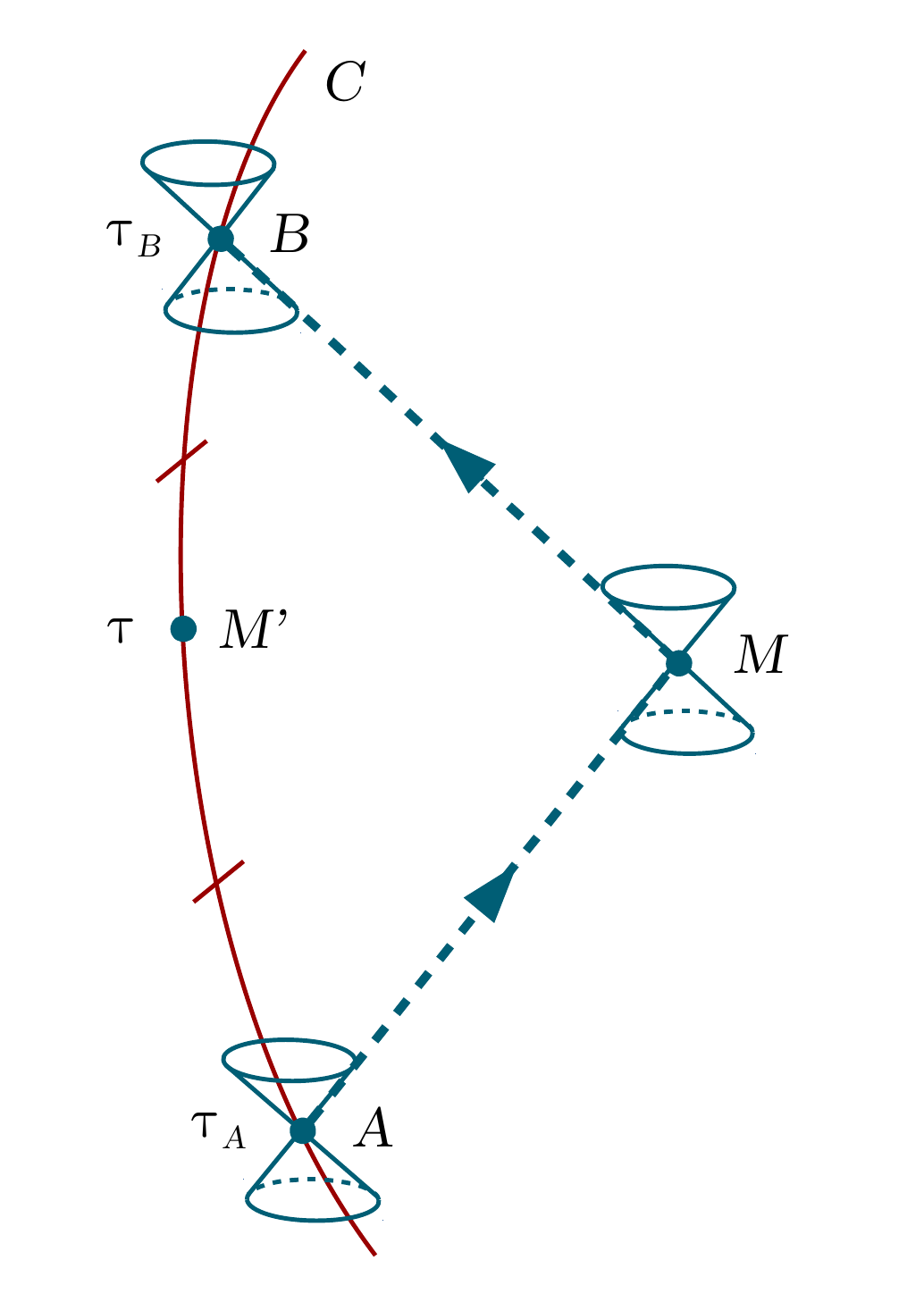}   
	   \end{center}
	\caption{\label{fig:einstein_sync} Illustration of the Einstein synchronization convention.}
   \end{minipage} \hfill
   \begin{minipage}[c]{.46\linewidth}
	\includegraphics[width=\linewidth]{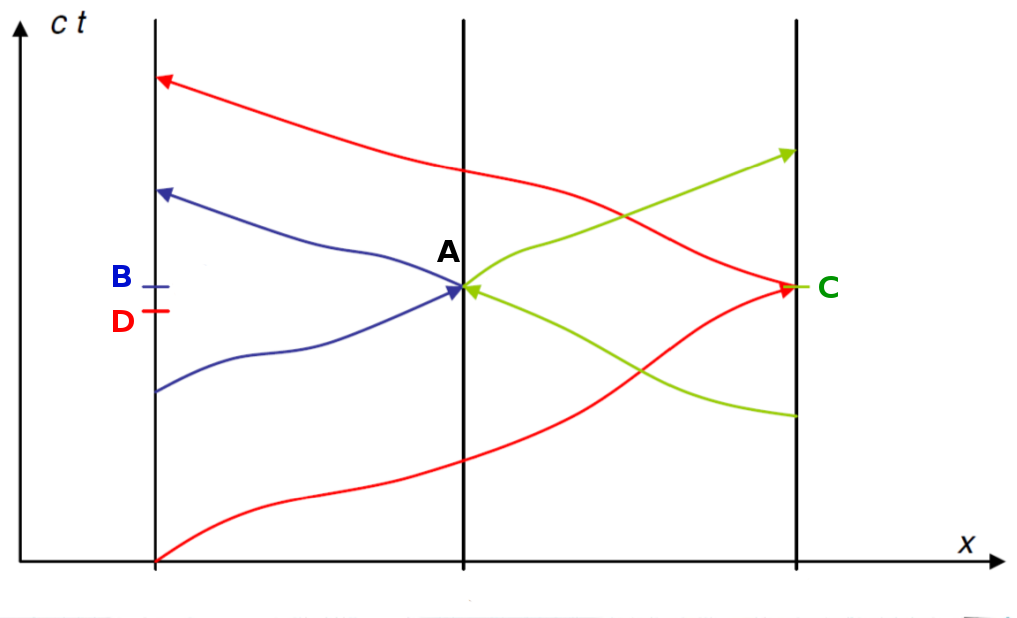}
   	\caption{\label{fig:einstein_trans} Einstein synchronization convention is not transitive: events $B$ and $C$ are defined as simultaneous with $A$ thanks to the convention, while $D$ is defined as simultaneous to $C$. However, events $B$ and $D$ do not coincide.}
   \end{minipage}
\end{figure}

Let us define an observer $\Obs$ with trajectory $\C$ parameterized by its proper time $\tau$, and an event $M$ which does not belong to $\C$ (see figure~\ref{fig:einstein_sync}). How can we define an event on $\C$ which is simultaneous with event $M$. This is possible in the Newtonian space which is Euclidean, and therefore time is absolute, i.e. independent from the observer. However in GR proper time is defined only along the worldline of the observer and is not a global property of space-time. Only the light cone is a fundamental element of a given space-time. The light cone is the collection of vectors $\w{v}\in\T_P(\M)$ which satisfies $\g{v}{v} = 0$. It is independent of the observer and divides the tangent space in three parts, \emph{past} and \emph{future} containing \emph{time vectors} which satisfy $\g{v}{v} < 0$, and a third part containing \emph{space vectors} which satisfy $\g{v}{v} > 0$. It is supposed here that the metric tensor has a signature $(-,+,+,+)$, i.e. at least one basis of the tangent space exists for which $\g{v}{w} = -v^0 w^0 + v^1 w^1 + v^2 w^2 + v^3 w^3$, where $\w{v},\w{w}\in\T_P(\M)$.

With the notion of the light cone, space-time can be time oriented, but it does not say which set of events can be considered simultaneous. Indeed in GR simultaneity can only be conventional and not an intrinsic property of space-time. Einstein has suggested an operational definition of simultaneity. Suppose that an observer $\Obs$ is equipped with a clock and a system to send and receive electromagnetic signals. A signal is sent at event $A\in\C$, received and reflected with no delay at event $M$ and finally received at event $B\in\C$ (see figure~\ref{fig:einstein_sync}). The proper times along $\C$ corresponding to events $A$ and $B$, respectively $\tau_A$ and $\tau_B$, are measured with the clock. By convention, the event $M'$ simultaneous to $M$ along the observer trajectory corresponds to proper time:
\begin{equation}
	\tau = \frac{1}{2} (\tau_A+\tau_B) = \tau_A + \frac{1}{2} (\tau_B - \tau_A) \ .
	\label{eq:einstein_sync}
\end{equation}
This convention is usually called \emph{Einstein synchronization}. It is a geometrical convention based on the concept of light cones, and an operational convention based on the exchange of electromagnetic signals. However this convention is not transitive. This is illustrated in figure~\ref{fig:einstein_trans}. Events $B$ and $C$ are simultaneous to $A$ (with Einstein synchronization); $D$ is simultaneous to $C$ but $D$ and $B$ events do not generally coincide. Therefore this convention is not practical to define global timescales such as the TAI (Temps Atomique International). This problem was discussed in~\cite{Cohen1977,Cohen1983,Cohen1984} in the context of satellite clock synchronization. However in these articles the problem was thought as ``synchronization errors''. But it was in fact well understood in the context of general relativity, as noted in~\cite{Podlaha1984,Ashby1984a,Ashby1985}.

\begin{figure}
   \begin{minipage}[c]{.46\linewidth}
	\includegraphics[width=\linewidth]{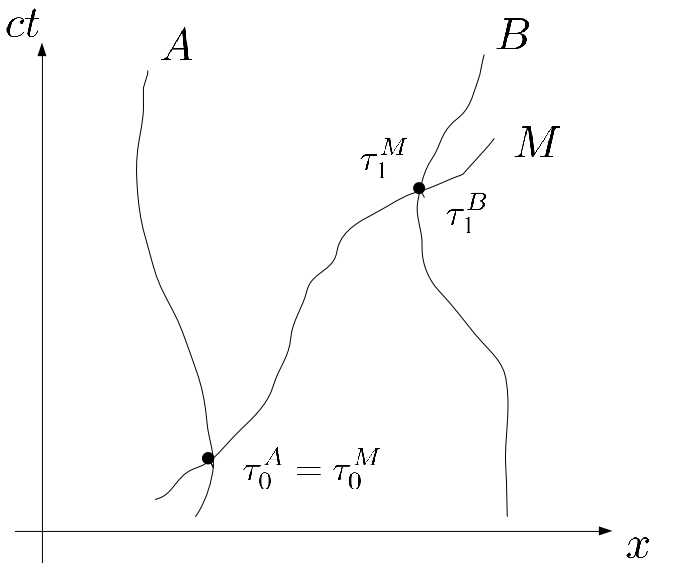}
	\caption{\label{fig:SCS} Slow clock transport synchronization convention.}
   \end{minipage} \hfill
   \begin{minipage}[c]{.46\linewidth}
	\includegraphics[width=0.8\linewidth]{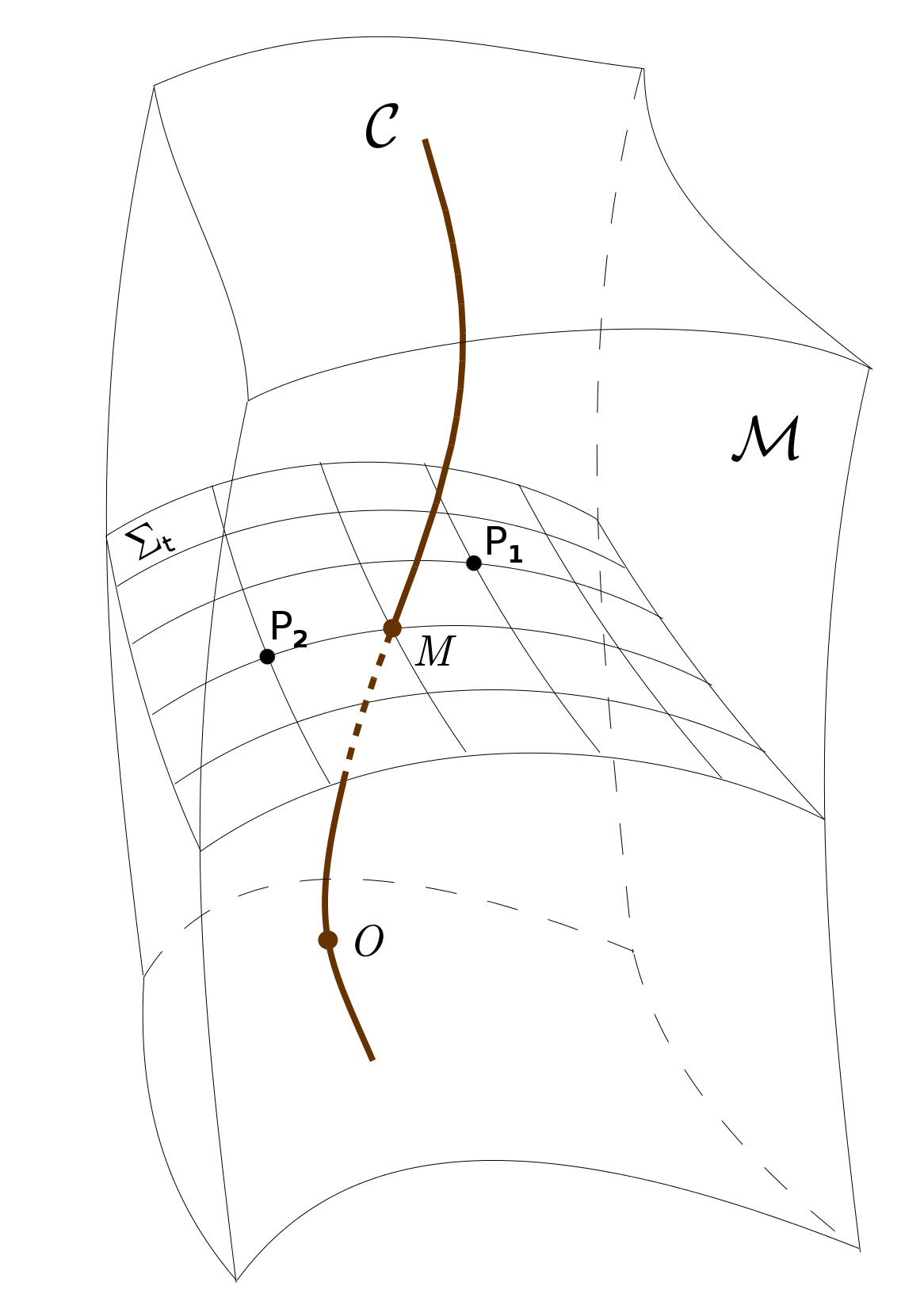}
   	\caption{\label{fig:hyper} Coordinate synchronization convention.}
   \end{minipage}
\end{figure}

Another convention is the slow clock transport synchronization. Let us define three clocks $A$, $B$ and $M$ with corresponding worldlines $\C_A$, $\C_B$ and $\C_M$ (see fig.\ref{fig:SCS}). Clocks $A$ and $M$ are compared locally at event $A_0$ such that clock $M$ proper time is set to $\tau^M = \tau^M_0 = \tau^A_0$ at this event. Then clock $M$ goes toward clock $B$ and crosses worldline $\C_B$ at event $B_1$, where $\tau^M = \tau^M_1$ and $\tau^B = \tau^B_1$. The event $B_0$ on $\C_B$ is defined simultaneous to event $A_0$ on $\C_A$ by the slow clock transport synchronization convention with:
\begin{equation}
	\tau_0^B = \lim_{v \rightarrow 0} [\tau_1^B - (\tau_1^M - \tau_0^M)] \ ,
	\label{eq:SCS}
\end{equation}
where $v$ is the coordinate velocity of clock $M$. The limit condition of null velocity is not feasible in a real experiment. Therefore, operationally, this convention depends on the particular trajectory of the mobile clock $M$, and reaches a different synchronization than the Einstein synchronization convention, as shown in~\cite{Cohen1983}. However, in special relativity, i.e. with a null background curvature, it can be shown that both synchronization conventions are equivalent. If the space-time geometry and the mobile clock trajectory are sufficiently known, then in the weak-field and low velocity approximation it is possible to use the coordinate time of clock $M$ at events $A_0$ and $B_1$ instead of its proper time, so that the convention will not depend on the particular trajectory of the mobile clock. However it will depend on the relativistic coordinate system chosen to calculate the coordinate time. The inaccuracy of the time transfer operated with this convention can be assessed with the closing relation: $\tau_0^A = \lim_{v \rightarrow 0} [\tau_1^A - (\tau_1^M - \tau_0^M)]$.

Finally, we define the coordinate synchronization convention: two events $P_1$ and $P_2$ of coordinates $\{x^\alpha_1\}$ and $\{x^\alpha_2\}$ are considered to be simultaneous if the values of their time coordinates are equal: $x^0_1 = x^0_2$ (see fig.\ref{fig:hyper}). This definition follows the definition of simultaneity adopted in special relativity in~\cite{Born1962,Moller1976}. It is convenient to introduce three-dimensional hypersurfaces with constant time coordinate $t$: $\Sigma_t \equiv \{P\in{\cal M},x^{0}_P=ct\}$. By choosing a particular relativistic reference system we introduce a conventional foliation of space-time, giving the hypersurfaces of simultaneity. The synchronization of clocks with this convention obviously depends on the chosen reference system. It is the most commonly used convention for the building of timescales such as TAI and GNSS timescales. Indeed, this convention is very similar to what is well known in Newtonian physics where the foliation of space-time is absolute. For this convention to become operational it is necessary to define conventional relativistic reference systems. In special relativity, for clocks which are at rest with respect to an inertial reference system, the Einstein synchronization is a convenient procedure to achieve coordinate synchronization of clocks.

It was proposed in~\cite{Ashby1979} to build a global ``coordinate time grid'' on and around the Earth, therefore realizing the idea of coordinate synchronization convention for clocks, without any problem of transitivity. The authors proposed to take as a reference a clock on the geoid\footnote{In the Newtonian sense, the geoid is the equipotential of the Earth's gravity (Newtonian) potential, which best coincides with the (mean) surface of the oceans.}, i.e. to choose a conventional reference system $\R$ such that the proper time of a clock at rest on the geoid coincides with its coordinate time in $\R$. We will see later that this choice is convenient because it implies a simple link between the relativistic correction of a clock, in order to realize the coordinate time synchronization, and its altitude. The authors in~\cite{Ashby1979} detailed several operational methods of time transfer using the coordinate synchronization convention: portable clocks, one-way and two-way synchronization with electromagnetic signals. The same authors in~\cite{Allan1986} estimated the main limitation on the determination of coordinate time: the knowledge of the geoid.

It is interesting to note that the question of synchronization of clocks in non-inertial reference systems raised a controversy in the 80's, driven by the development of GPS and the need for a global timescale on Earth. This is reviewed in~\cite{Skalafuris1985}, where the author concludes: ``In principle, the curved Schwarzschild space cannot be imbedded in a four-dimensional flat space without the addition of more dimensions. Thus the theoretical basis for the GPS navigational scheme would appear to be flawed, and a new algorithm would have to be constructed''. Indeed, coordinate time synchronization can only be theoretically realized in approximation schemes, e.g. post-Newtonian approximation as reviewed in~\cite{Ashby2003} for GPS. A different relativistic approach to this problem has been initiated in~\cite{Coll2003}, where the idea is to give to a constellation of satellites the possibility to constitute by itself a primary and autonomous positioning system, without any need for synchronization of the clocks. Such a \emph{relativistic positioning system} is defined with the introduction of \emph{emission coordinates}, which have been re-introduced by several authors in the context of navigation systems~\cite{Rovelli2002,Lachieze-Rey2006,Coll1991,Blagojevic2002,Delva2011,Bini2008,Tartaglia2010,Puchades2012,Bunandar2011}.

A resolution concerning the global ``coordinate time grid'' was proposed by N.~Ashby at the International Astronomical Union (IAU) Symposium No.~114, reported in~\cite{Brumberg1986}:
\begin{enumerate}
	\item To adopt the coordinate time system (as approved by the international committees CCDC and CCIR) as a global time scale for the Earth;
	\item To continue further investigations for the determination and adjustment of the International Atomic Time (TAI) and the Terrestrial Dynamic Time (TDT).
\end{enumerate}
The resolution was not adopted, but the chairman of the Scientific Organizing Committee, J.~Kovalesky, considered that specialists in Celestial Mechanics and Astrometry needed more time to study the problem in competent commissions of IAU. Following this Symposium, several authors have contributed to the definition of global coordinate times~\cite{Guinot1986,Guinot1988,Huang1989,Brumberg1990,Klioner1992}. 

As the definition of coordinate timescales necessitates the definition of a four dimensional relativistic reference system, the IAU working group had several complementary tasks in hand (Resolution C2 of the IAU General Assembly in 1985\footnote{All IAU Resolutions can be found at \url{http://www.iau.org/administration/resolutions/general_assemblies/}}):
\begin{enumerate}
	\item the definition of the Conventional Terrestrial and Conventional Celestial Reference Systems,
	\item ways of specifying practical realizations of these systems,
	\item methods of determining the relationships between these realizations, and
	\item a revision of the definitions of dynamical and atomic time to ensure their consistency with appropriate relativistic theories
\end{enumerate}
Moreover, the President of the International Association of Geodesy (IAG) was invited to ``appoint a representative to the working group for appropriate coordination on matters relevant to Geodesy''. This work eventually led to the set of IAU Resolutions in 1991 and 2000 that define the present reference systems.

\subsection{Relativistic reference systems}
\label{sec:ref_syst}
Several approaches have been considered for the definition of relativistic reference systems. Generalised Fermi coordinates were considered in~\cite{Ashby1984,Fukushima1988,Ashby1986}. However, the use of Fermi coordinates is not adapted to self-gravitating bodies for which mass-energy contributes to the determination of the initial metric $\bm{g}$ when solving the Einstein equations. For this reason, harmonic coordinates are preferred and recommended for the definition and realization of relativistic celestial reference systems~\cite{Kopejkin1988,Soffel1989,Brumberg1989,Brumberg1989a,Damour1991}, where the frame origin can be centered on the center-of-mass of a massive body. One drawback of harmonic frames is that the harmonic gauge condition does not admit rigidly rotating frames~\cite[chapter~8]{Kopeikin2011}. Other recent approaches are based on a perturbed Schwarzschild metric~\cite{Kostic2015}, or on the Kerr metric~\cite{Bini2016} in the different context of a slowly rotating astronomical object.

Following the pioneering works, a set of Resolutions was adopted at the IAU General Assembly in Manchester in the year 2000~\cite{Soffel2003}:
\begin{itemize}
	\item B1.3: definition of the Barycentric and Geocentric Celestial Reference Systems (BCRS and GCRS)
	\item B1.4: form of the Earth post-Newtonian potential expansion
	\item B1.5: time transformations and realization of coordinate times in the Solar System (uncertainty $< 5 \times 10^{-18}$ in rate and 0.2 ps in phase amplitude for locations farther than a few Solar radii from the Sun)
	\item B1.9: definition of Terrestrial Time (TT) 
\end{itemize}

We summarize here very briefly these resolutions. A relativistic reference system is implicitely defined by giving the components of the metric tensor in this reference system, in addition to a conventional spatial origin and orientation for the spatial part of the frame, and a conventional time origin for the time coordinate (the time orientation is trivial). The metric tensor is a solution of the Einstein equations in the low velocity and weak gravitational field approximation, for an ensemble of $N$ bodies.

The Solar System Barycentric Celestial Reference System (BCRS), recommended by the IAU Resolutions, can be used to model light propagation from distant celestial objects and the motion of bodies within the Solar System. It is defined with:
\begin{eqnarray}
	g_{00} &= -1 + \frac{2 w(\w{x})}{c^2}  - \frac{2 w(\w{x})^2}{c^4} \label{eq:BCRS1} \ ,\\
	g_{0i} &=  -\frac{4}{c^3} w^i (\w{x}) \ , \label{eq:BCRS2}\\
	g_{ij} & = \delta_{ij} \left(1 + \frac{2 w(\w{x})}{c^2} \right) \ , \label{eq:BCRS3}
\end{eqnarray}
where $\w{x}\equiv \{ ct,x^i \}$, with $i=1\dots3$, $w$ and $w^i$ are scalar and vector potentials. Its origin is at the barycenter of the Solar System masses, while the orientation of spatial axes is fixed up to a constant time-independent rotation matrix about the origin (a natural choice is the ICRS orientation which is fixed w.r.t. distant quasars). The coordinate time $t$ is called Barycentric Coordinate Time (TCB). The origin of TCB is defined w.r.t. TAI: its value on 1977 January 1, 00:00:00 TAI (JD = 2,443,144.5 TAI) must be 1977 January 1, 00:00:32.184. 

The unit of measurement of TCB should be chosen so that it is consistent with the SI second. An interesting discussion about timescales units can be found in~\cite{Klioner2009}. As coordinate times such as TCB are not proper times, they cannot be directly measured by clocks. They are calculated using the corresponding metric components, e.g. eqs.(\ref{eq:BCRS1})-(\ref{eq:BCRS3}) for TCB, in combination with eq.(\ref{eq:tau_t}), which has to be inverted. Indeed, the basic observables to build timescales are the readings of proper times on clocks, which are local experiments. If the clocks are realizing the SI second, then the timescales calculated from these measurements are also in SI units, and the unit of such time coordinate could be named ``SI-induced second''.

The second relativistic reference system, recommended by the IAU Resolutions, is the Geocentric Celestial Reference System (GCRS). It can be used to model phenomenon in the vicinity of the Earth, such as its gravity field, artificial satellites orbiting the Earth or Earth rotation. It is defined with:
\begin{eqnarray}
	G_{00} &= -1 + \frac{2 V (\w{X})}{c^2}  - \frac{2 V(\w{X})^2}{c^4} \ , \label{eq:GCRS1}\\
	G_{0i} &=  -\frac{4}{c^3} V^i (\w{X}) \ , \label{eq:GCRS2}\\
	G_{ij} & = \delta_{ij} \left(1 + \frac{2 V(\w{X})}{c^2} \right) \ , \label{eq:GCRS3}
\end{eqnarray}
where $(\w{X})\equiv \{ cT,X^i \}$, and $V$ and $V^i$ are scalar and vector potentials. Note that we use notation $V$ instead of usual notation $W$ because $W$ is commonly used in geodesy for the gravity potential. The frame origin is at the centre of mass of the Earth, and the orientation of spatial axes is fixed w.r.t the spatial part of the BCRS. The coordinate time $T$ is called Geocentric Coordinate Time (TCG). It has the same origin and unit as TCB.

TCG is the proper time of a clock at infinity, and is not convenient because its rate differs from the one of clocks on the ground. Therefore IAU Resolutions introduced Terrestrial Time (TT), which differs from TCG by a constant rate $L_G=6.969290134\times10^{-10}$:
\begin{equation}\label{dTTdTCG}
	\frac{\D ({\rm TT})}{\D ({\rm TCG})} = 1 - L_G.
\end{equation}
The origin of TT is defined so that TCG coincides with TT in origin: TT = TAI + 32.184~s on 1977 January $1^{st}$, 0 h TAI. TT is a theoretical timescale and can have different realizations, e.g. TT(BIPM), or TT(TAI) = TAI + 32.184 s. (see e.g.~\cite{Arias2011}).

\subsection{Chronometric geodesy}
\label{sec:chrono}
Chronometric geodesy is the use of clocks to determine the space-time metric. Indeed, the gravitational redshift effect discovered by Einstein must be taken into account when comparing the frequencies of distant clocks. Instead of using our knowledge of the Earth's gravitational field to predict frequency shifts between distant clocks, one can revert the problem and ask if the measurement of frequency shifts between distant clocks can improve our knowledge of the gravitational field. To do simple orders of magnitude estimates it is good to have in mind some correspondences:
\begin{equation}
1 \ {\rm meter} \leftrightarrow \frac{\Delta \nu}{\nu} \sim 10^{-16} \leftrightarrow \Delta W \sim 10 \ {\rm m}^2 \ {\rm s}^{-2} \ ,
\label{eq:corr0}
\end{equation}
where 1~meter is the height difference between two clocks, $\Delta \nu$ is the frequency difference in a frequency transfer between the same two clocks, and $\Delta W$ is the gravity potential difference between the locations of these clocks.

From this correspondence, we can already recognize two direct applications of clocks in geodesy: if we are capable of comparing clocks to $10^{-16}$ accuracy, we can determine height differences between clocks with one meter accuracy (levelling), or determine geopotential differences with 10~m$^2$~s$^{-2}$ accuracy.

To the knowledge of the authors, the latter technique was first mentioned in the geodetic literature by Bjerhammar~\cite{Bjerhammar1975} within a short section on a ``new physical geodesy''. Vermeer~\cite{Vermeer1983} introduced the term ``chronometric levelling'', while Bjerhammar~\cite{Bjerhammar1985} discussed the clock-based levelling approach under the title ``relativistic geodesy'', and also included a definition of a relativistic geoid.
The term ``chronometric'' seems well suited for qualifying the method of using clocks to determine directly gravity potential differences, as ``chronometry'' is the science of the measurement of time. However the term ``levelling'' seems to be too restrictive with respect to all the applications one  could think of using the results of clock comparisons. Therefore we will use the  term ``chronometric geodesy'' to name the scientific discipline that deals with  the measurement and representation of the Earth, including its gravity  field, with the help of atomic clocks.  It is sometimes also named ``clock-based  geodesy'', or ``relativistic geodesy''. However this last designation is  improper as relativistic geodesy aims at describing all possible techniques  (including e.g. gravimetry, gradiometry, VLBI, Earth rotation, \ldots) in a relativistic framework. The natural arena of chronometric geodesy is the four-dimensional space-time. At the lowest order, there is proportionality between relative frequency shift  measurements -- corrected from the first order Doppler effect -- and (Newtonian)  gravity potential differences. To calculate this relation one does not need the theory of general relativity, but only to  postulate Local Position Invariance. Therefore, if the measurement accuracy does not reach the magnitude of the  higher order terms, it is perfectly possible to use clock comparison measurements -- corrected for the first order Doppler effect  -- as a direct measurement of (differences of) the gravity potential that is considered in classical geodesy. Comparisons between two clocks on the ground generally use a third reference clock in space, or an optical fibre on the ground (see section~\ref{sec:tech}). 

In his article, Martin Vermeer explores the ``possibilities for technical realisation of a system for measuring potential differences over intercontinental distances'' using clock comparisons~\cite{Vermeer1983}. The two main ingredients are, of course, accurate clocks and a mean to compare them. He considers hydrogen maser clocks. For the links he considers a 2-way satellite link over a geostationary satellite, or GPS receivers in interferometric mode. He has also considered a way to compare proper frequencies of the different hydrogen maser clocks.  Today this can be overcome by comparing primary frequency standards (PFS, see section~\ref{sec:AFS}), which have a well defined proper frequency based on the transition of Caesium~133 used for the definition of the second. Secondary frequency standards (SFS), i.e. standards based on a transition other than the defining one, may also be used if the uncertainty in systematic effects has been fully evaluated, and the frequency measured against PFS. 

With the advent of optical clocks, it often happens that the evaluation of systematics can be done more accurately than for PFS. This was one of the purpose of the European project\footnote{\url{projects.npl.co.uk/itoc}} of ``International timescales with optical clocks''~\cite{Margolis2013}, where optical clocks based on different atoms are compared to each other locally, and to PFS. Within this project, a proof-of-principle experiment of chronometric geodesy was done by comparing two optical clocks separated by a height difference of around 1000~m, using an optical fibre link~\cite{Grotti2017}. 

Few authors have seriously considered chronometric geodesy in the past. Following Vermeer's idea, the possibility of using GPS observations to solve the problem of determinating the geoid heights has been explored in~\cite{Brumberg2002}. The authors considered two techniques based on frequency comparisons and direct clock readings. However, they leave aside the practical feasibility of such techniques. The value and future applicability of chronometric geodesy has been discussed in~\cite{Bondarescu2012}, including direct geoid mapping on continents and joint gravity-geopotential surveying to invert for subsurface density anomalies. They find that a geoid perturbation caused by a 1.5 km radius sphere with 20 percent density anomaly buried at 2 km depth in the Earth's crust is already detectable by atomic clocks with an achievable accuracy of $10^{-18}$. The  potentiality of the new generation of atomic clocks has been shown in~\cite{Chou2010}, based on optical transitions, to measure heights with a resolution of around 30~cm. The possibility of determining the geopotential at high spatial resolution thanks to chronometric geodesy is thoroughly explored and evaluated in~\cite{Lion2017}. This work will be detailed in section~\ref{sec:high_res}.

\subsection{The chronometric geoid}
\label{sec:chronogeoid}
Arne Bjerhammar in 1985 gives a precise definition of the ``relativistic geoid''~\cite{Bjerhammar1985, Bjerhammar1986}:  \begin{quote} ``The relativistic geoid is the surface where precise clocks run with the same  speed and the surface is nearest to mean sea level'' \end{quote} This is an operational definition, which has been translated in the context of post-Newtonian theory~\cite{Soffel1988,Soffel1989}. In these two articles a different operational definition of the relativistic geoid has been introduced based on gravimetric measurements: a surface orthogonal everywhere to the direction of the plumb-line and closest to mean sea level. The authors call the two surfaces obtained with clocks and gravimetric measurements respectively the ``u-geoid'' and the ``a-geoid''. They have shown that these two surfaces coincide in the case of a stationary metric. In order to distinguish the operational definition  of the geoid from its theoretical description, it is less ambiguous to give a name based  on the particular technique to measure it. The term ``relativistic geoid'' is too vague as  Soffel et al.~\cite{Soffel1988} have defined two different ones. The names chosen by Soffel et al.~are not particularly explicit, so instead of ``u-geoid'' and ``a-geoid'' one can call them ``chronometric geoid'' and ``gravimetric geoid'' respectively. There can be no confusion with the geoid derived from satellite measurements, as this is a  quasi-geoid that does not coincide with the geoid on the continents~\cite{Hofmann-Wellenhof2006}.  Other considerations on the chronometric geoid can be found in~\cite{Kopejkin1991, Kopeikin2011, Muller2007}.

We notice that the problem of defining a reference isochronometric surface is closely related to the problem of realizing Terrestrial Time (TT). This is developed in more details in section~\ref{sec:TT}.

Recently, extensive work has been done aiming at developing an exact relativistic theory for the Earth's geoid undulation~\cite{Kopeikin2015}, as well as developing a theory for the reference level surface in the context of post-newtonian gravity~\cite{Kopeikin2016,Kopeikin2016a}. This goes beyond the problem of the realization of a reference isochronometric surface and tackles the tough work of extending all concepts of classical physical geodesy (see e.g.~\cite{Hofmann-Wellenhof2006}) in the framework of general relativity.


\section{Comparisons of frequency standards}
\label{sec:frequ_comp}
	
\subsection{The Einstein Equivalence Principle}

\begin{figure}[!ht]
   \begin{center}
      \includegraphics[width=\linewidth]{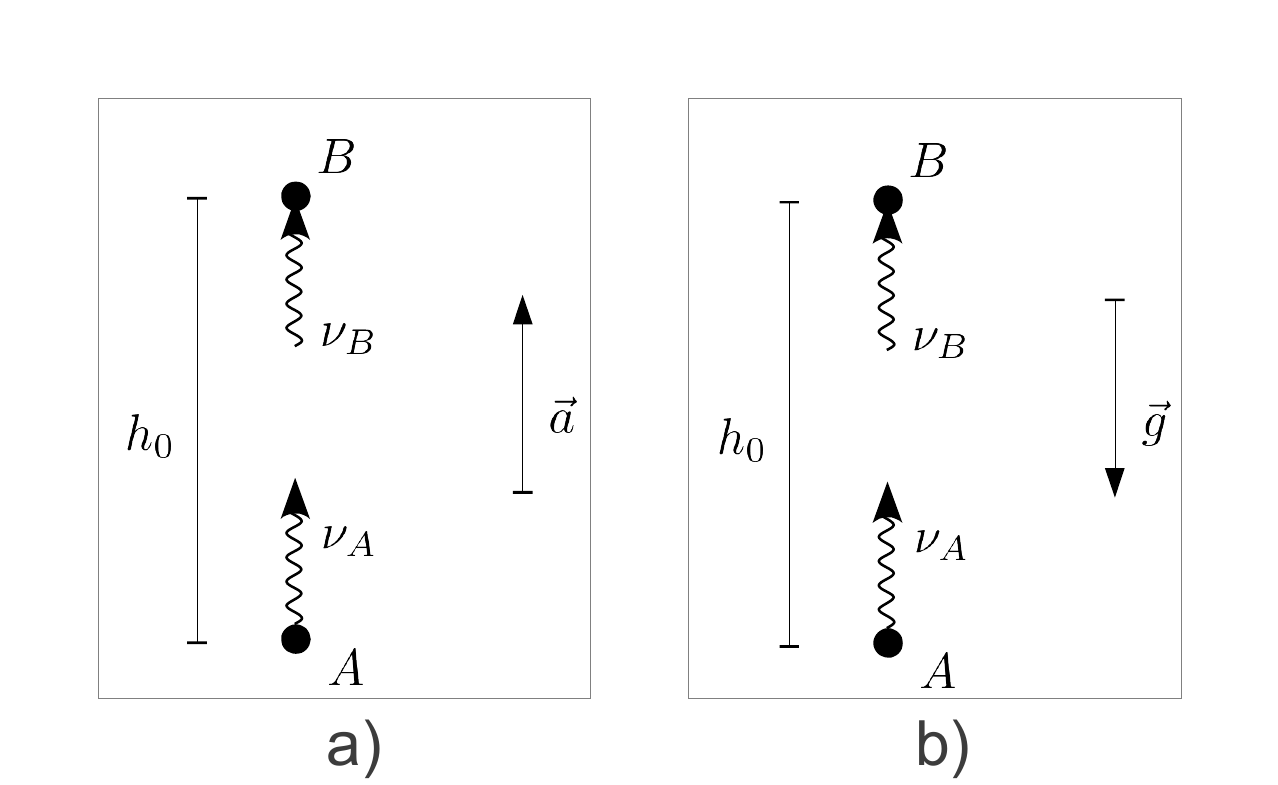}
   \end{center}
   \caption{\label{fig:red} \footnotesize A photon of frequency $\nu_A$  is emitted at point A  toward point B, where the measured
   frequency is $\nu_B$. a) $A$ and $B$ are  two points at rest in an accelerated frame, with acceleration $\vec{a}$ in  the same
   direction as the emitted photon. b) $A$ and $B$ are at rest in a non  accelerated (locally inertial) frame in presence of a
   gravitational field such  that $\vec{g} = - \vec{a}$.}
\end{figure}

Let's consider a photon emitted at a point $A$ in an accelerated reference system,
toward a point $B$ which lies in the direction of the acceleration~(see fig.\ref{fig:red}).
We assume that both  points are separated by a distance $h_0$, as measured in the accelerated frame. The photon time of flight is $\delta t = h_0/c$, and the frame velocity during this time increases by $\delta v = a \delta t = ah_0/c$, where $a$ is the magnitude of the frame acceleration ${\vec a}$. The frequency at point $B$ (reception) is then shifted because of the Doppler effect, compared to the frequency at point $A$ (emission), by an amount:

\begin{equation}
\frac{\nu_B}{\nu_A} = 1 - \frac{\delta v}{c} = 1 - \frac{ah_0}{c^2} \ .
\end{equation}

Now, the Einstein Equivalence Principle (EEP) postulates that a gravitational field ${\vec g}$ is locally equivalent to an acceleration field ${\vec a} = - {\vec g}$. We deduce that in a  non accelerated (locally inertial) frame in presence of a gravitational field ${\vec g}$:
\begin{equation}
\frac{\nu_B}{\nu_A} = 1 - \frac{gh_0}{c^2} \ ,
\end{equation}
where $g=|{\vec g}|$, $\nu_A$ is the photon frequency at emission (strong  gravitational potential) and $\nu_B$ is the photon frequency
at reception (weak  gravitational potential). As $\nu_B<\nu_A$, it is usual to say that the frequency at  the point of reception is ``red-shifted''.
One can consider it in terms of  conservation of energy. Intuitively, the photon that goes from A to B has to  ``work'' to be able
to escape the gravitational field, then it looses energy and  its frequency decreases by virtue of $E= h \nu$, with $h$ the Planck constant.

If two ideal clocks are placed in $A$ and $B$ and the clock at $A$ (strong gravitational potential) is used to generate the signal $\nu_A$, then the signal received at $B$ (weak gravitational potential) has a lower frequency than a signal locally generated by the clock at $B$.

\subsection{Relativistic frequency transfer}
\label{sec:AFS}
Let's consider two atomic frequency standards (AFS) $A$ and $B$ which deliver the proper frequencies $f_A$ and $f_B$. These two frequencies can be different if the two AFS are based on different atom transitions. Following the Bureau International des Poids et Mesures~(BIPM) we name primary frequency standards (PFS) the AFS based on the atom of Caesium 133, more commonly named Caesium Fountains. The best PFS have a very low relative accuracy in the range $10^{-15}$ -- $10^{-16}$~(see e.g.~\cite{Guena2017}). Then, we name secondary frequency standards (SFS) the AFS which are based on a different atom than the Caesium atoms. The CCL-CCTF frequency standards working group is in charge of producing and maintaining a single list of recommended values of standard frequencies for the practical realization of secondary representations of the second\footnote{see \url{http://www.bipm.org/en/publications/mises-en-pratique/standard-frequencies.html}}. SFS can have a relative accuracy down to the range $10^{-17}$ -- $10^{-18}$~\cite{Lisdat2016,Schioppo2017,Huntemann2016}. See also~\cite{Margolis2015a,Margolis2016}, where a method is presented for analysing over-determined sets of clock frequency comparison data involving standards based on a number of different reference transitions.

The goal of a frequency comparison between two AFS $A$ and $B$ is to determine the ratio of their frequency $f_A/f_B$. The most used technique for frequency comparison nowadays is the transmission of an electromagnetic signal between $A$ and $B$, reaching the following formula:
\begin{equation}
	\frac{f_A}{f_B} = \frac{f_A}{\nu_A}\frac{\nu_A}{\nu_B}\frac{\nu_B}{f_B} \ ,
	\label{eq:ratioC}
\end{equation}
where $\nu_A$ is proper frequency of the photon at the time of emission $t_A$, and $\nu_B$ is proper frequency of the same photon at the time of reception $t_B$. The ratio $\nu_A/f_A$ is known or measured, $\nu_B/f_B$ is measured while $\nu_A/\nu_B$ has to be modelled and calculated. 

Let $S(x^\alpha)$ be the phase of the
electromagnetic signal emitted by clock $A$.  It can be shown that light rays
are contained in hypersurfaces of constant phase. The frequency measured by
$A/B$ is:
\begin{equation}
\label{eq:nudef}
\nu_{A/B} = \frac{1}{2\pi} \frac{\dd S}{\dd \tau_{A/B}} \ ,
\end{equation}
where $\tau_{A/B}$ is the proper time along the worldline of clock $A/B$. We
introduce the wave vector $k^{A/B}_\alpha = (\partial_\alpha S)_{A/B}$ to obtain:
\begin{equation}
\nu_{A/B}= \frac{1}{2\pi} k^{A/B}_\alpha u^\alpha_{A/B} \ ,
\end{equation}
where $u^\alpha_{A/B} = \dd x_{A/B}^\alpha / \dd \tau$ is the four-velocity of
clock $A/B$.
Finally, we obtain a fundamental relation for frequency transfer:
\begin{equation}
\frac{\nu_A}{\nu_B} = \frac{k^A_\alpha u^\alpha_A}{k^B_\alpha u^\alpha_B} \ .
\end{equation}
This formula does not depend on a particular theory, and then can be used to
perform tests of general relativity. It is needed in the context of chronometric
geodesy in order to calculate the gravity potential difference between two clocks for which the ratio $f_A/f_B$ is well known.

Introducing $v^i = \dd x^i / \dd t$ and
$\hat k_i = k_i / k_0$, it is usually written as:
\begin{equation}
\label{eq:nuk}
\frac{\nu_A}{\nu_B} = \frac{u^0_A}{u^0_B} \frac{k_0^A}{k_0^B} \frac{ 1 + \frac{\hat
k^A_i v^i_A}{c} }{ 1 + \frac{\hat k^B_i v^i_B}{c} } \ .
\end{equation}
From eq.~(\ref{eq:nudef}) we deduce that:
\begin{equation}
	\frac{\nu_A}{\nu_B} = \frac{\D\tau_B}{\D\tau_A} = \left( \frac{\D \tau}{\D t} \right)_A^{-1} \left( \frac{\D \tau}{\D t} \right)_B \frac{\D t_B}{\D t_A} \ .
	\label{eq:ratioEM}
\end{equation}
The derivative $(\D t_B/\D t_A)$ is affected by processes in the frequency transfer itself and depends on the particular technique used for the frequency comparison. It is considered in more details in section~\ref{sec:tech}.

The derivatives $(\D\tau/\D t)$ in~(\ref{eq:ratioEM}) do not depend on the frequency transfer technique but just on the state (velocity and location) of the emitting and receiving AFS. In sections~\ref{sec:geod} and~\ref{sec:itoc} we focus on the best practical determination of these terms. Indeed, calculation of these terms is limited in accuracy by our knowledge of the Earth's gravitational field. We note that quantity $(\nu_A/\nu_B)$ in~(\ref{eq:ratioEM}) is a scalar, therefore invariant under general coordinate transformation (see section~\ref{sec:observers}). However the splitting of this quantity as written on the r.h.s. of~(\ref{eq:ratioEM}) is not invariant and depends on the particular relativistic reference system used for the splitting. The choice of a particular relativistic reference system gives a conventional meaning to simultaneity: two events are simultaneous if they have the same time coordinate $t$ (see section~\ref{sec:sync}) for free and guided propagation.

For applications on the Earth, such as (ground) clock syntonization (section~\ref{sec:synto}) and the realization of a worldwide coordinate time (section~\ref{sec:TT}), a natural choice of a relativistic reference system is the spatial part of the geocentric celestial reference system (GCRS) together with the terrestrial time (TT) as a coordinate time (see section~\ref{sec:ref_syst}). Following~\cite{Wolf1995,Denker2017}, the coordinate to proper time transformation can be written down to a relative accuracy of $10^{-18}$ as:
\begin{equation}
	\frac{\D\tau}{\D TT} =  1 + L_G -\frac{1}{c^2}\left[ W^{\rm{static}} + W^{\rm{temp}} \right] \ . 
	\label{eq:transfoTT}
\end{equation}
where $L_G$ is a constant defined in section~\ref{sec:ref_syst}, and $W=V+Z$ is the gravity (gravitational plus centrifugal) potential, commonly used in geodesy (see e.g.~\cite{Hofmann-Wellenhof2006,Torge1991}). The gravity potential is split into a static part $W^{\rm{static}}$ and a part varying with time $W^{\rm{temp}}$. Neglected terms in Eq.~(\ref{eq:transfoTT}) are terms in $c^{-4}$ or smaller as well as one term of order $c^{-2}$ resulting from the coupling of higher order multipole moments of the Earth to the external tidal gravitational field. All the neglected terms in the transformation~(\ref{eq:transfoTT}) amount in the vicinity of the Earth to a few parts in $10^{-19}$ or less~\cite{Wolf1995}. 

The static part of the gravity potential $W^{\rm{static}}$ can be derived from geometric levelling or GNSS positions combined with a gravimetric geoid model. For instance, the best unified evaluation of the static gravity potential for several AFS in Europe was one of the main purposes of the ITOC project (see section~\ref{sec:itoc}). 

\subsection{Frequency transfer techniques}
\label{sec:tech}

We discuss in this section the foundations of two frequency transfer techniques widely used, based on the propagation of an electromagnetic signal either in free space or in an optical fiber. Free and guided propagation lead to different theoretical modelling approaches of the frequency transfer. We limit the presented results to one-way transfer and give appropriate references for two-way transfer techniques.

Free space time and frequency transfer can be realized using radiofrequency signals (of order 1-10~GHz) with well established techniques~\cite{Bauch2015}, and in the optical domain with lasers~\cite{Samain2015}. GNSS (Global Navigation Satellite Systems) \cite{Petit2015,Droste2015,Leute2016a,Dube2017,Baynham2018} and TWSTFT (Two-Way Satellite Time and Frequency Transfer) \cite{Kirchner1991,Piester2008,Bauch2015,Fujieda2016,Hachisu2014} have been widely used for years to perform clock comparisons and establish international timescales such as TAI~\cite{Arias2011}. The ACES MWL (MicroWave Link)~\cite{Meynadier2018} is being developed in the frame of the ACES (Atomic Clocks Ensemble in Space) experiment~\cite{Cacciapuoti2009a,Laurent2015}. New techniques using two-way laser links have been developed and operated, such as T2L2 (Time Transfer by Laser Light) \cite{Exertier2016,Rovera2016,Samain2014,Samain2015a,Exertier2014}, and others are in development, such as ELT \cite{Kodet2011,Prochazka2016}, which is part of the ACES experiment.

Existing free space frequency transfer techniques are in the range $10^{-15}$--$10^{-16}$ for the fractional frequency accuracy and stability, with the goal of being in the $10^{-17}$ range for the ACES experiment. However, they are not sufficient for the comparisons of optical clocks, which have fractional frequency accuracy and stability in the $10^{-17}$--$10^{-19}$ range~\cite{Campbell2017,Koller2017,Schioppo2017,Huntemann2016,Tyumenev2016,Lodewyck2016}. Therefore, phase-coherent optical links have been developped using principally an optical fibre as a medium for the propagation~\cite{Predehl2012,Lopez2012,Lisdat2016,Guena2017}, attaining spectacular stability and accuracy in the range $10^{-19}$ and below. However, phase coherent free space optical links are being developed~\cite{Chiodo2013,Djerroud2010,Giorgetta2013,Deschenes2016}. It is not clear yet if these techniques will be able to be as good as optical fibre techniques, mainly because of the effect of atmospheric turbulence~\cite{Sinclair2014,Sinclair2016,Robert2016}.

\subsubsection{Free space propagation comparisons}

In the case of propagation in free space, if we suppose that the space-time is stationary, i.e.~$\partial_0 g_{\alpha \beta} =
0$, then it can be shown that $k_0$ is constant along the light ray, meaning
that $k_0^A = k_0^B$. Then, from eq.~(\ref{eq:nuk}) and~(\ref{eq:ratioEM}) we
deduce that
\begin{equation}
\label{eq:dtAdtB}
\frac{\dd t_B}{\dd t_A} = \frac{ 1 + \frac{\hat
k^A_i v^i_A}{c} }{ 1 + \frac{\hat k^B_i v^i_B}{c} } \ .
\end{equation}

The quantity $\dd t_B / \dd t_A$ in eq.(\ref{eq:ratioEM}) can be computed with
several methods. Two different approaches are presented in some detail in Appendix A
of~\cite{Blanchet2001}: a direct integration of the null geodesic
equations, and a simpler way which is the differentiation of the time transfer
function. This second method is quite powerful: a general method has been
developed to calculate the time transfer function as a Post-Minkowskian (PM)
series, up to any order in $G$, the gravitational constant~\cite{Poncin-Lafitte2004,
Teyssandier2008}. See for example~\cite{Hees2012} for the calculation of the
one-way frequency shift up to the 2-PM approximation. This
method does not require the integration of the null geodesic equations. The
frequency shift is expressed as integral of functions defined from the metric
and its derivatives, and performed along a Minkowskian straight line.

Let A be the emitting station, with GCRS position $\vec{x}_A(t)$, and B the receiving station, with position $\vec{x}_B(t)$.
We use $t = \mathrm{TCG}$ and the calculated coordinate time intervals are in $TCG$. The corresponding time intervals in TT are obtained
by multiplying with $(1-L_G)$. We denote by $t_A$ the coordinate time at the instant of emission of an electromagnetic signal, and by $t_B$ the coordinate time at the instant of reception. We define $r_A=|\vec{x}_A(t_A)|$, $r_B=|\vec{x}_B(t_B)|$ and ${\rm R}_{AB}=|\vec{x}_B(t_B)-\vec{x}_A(t_A)|$, as well as the coordinate velocities $\vec{v}_A = \dd \vec{x}_A / \dd t (t_A)$ and $\vec{v}_B = \dd \vec{x}_B / \dd t (t_B)$. Then the frequency ratio can be expressed as~\cite{Blanchet2001}:
\begin{equation}
\label{eq:FS}
\frac{\nu_A}{\nu_B} = \frac
{1-{1\over c^2}\left[{{v}_B^2\over 2}+U_E(\vec{x}_B)\right]}
{1-{1\over c^2}\left[{{v}_A^2\over 2}+U_E(\vec{x}_A)\right]}
\frac{q_A }{q_B } \ ,
\end{equation}
where $U_E$ is the Newtonian potential of the Earth, and, if the desired accuracy is greater than $5 \times 10^{-17}$,
\begin{eqnarray}
\label{eq:q1}
q_A &= 1 - \frac{\vec{N}_{AB}\cdot\vec{v}_A}{c} - \frac{4GM_E}{c^3}
\frac{{\rm R}_{AB} \vec{N}_A \cdot \vec{v}_A + (r_A + r_B) \vec{N}_{AB} \cdot
\vec{v}_A}{ (r_A+r_B)^2 - {\rm R}_{AB}^2} \ , \\
\label{eq:q2}
q_B &= 1 - \frac{\vec{N}_{AB} \cdot \vec{v}_B}{c} - \frac{4GM_E}{c^3}
\frac{{\rm R}_{AB} \vec{N}_A \cdot \vec{v}_B - (r_A + r_B) \vec{N}_{AB} \cdot
\vec{v}_B}{ (r_A+r_B)^2 - {\rm R}_{AB}^2} \ ,
\end{eqnarray}
where $\vec{N}_{AB} = (\vec{x}_B(t_B)-\vec{x}_A(t_A)) / R_{AB}$, $G$ is the gravitational constant and $M_E$ is the mass of the Earth.

Note that formulas~(\ref{eq:q1}) and~(\ref{eq:q2}) have been obtained by assuming
that the field of the Earth is spherically symmetric.
If an accuracy lower than $5 \times 10^{-17}$ is required, it is necessary to take into account the $J_2$ terms in the Newtonian potential.

The terms of order $c^{-1}$ correspond to the relative Doppler effect between the clocks. Terms of order $c^{-2}$ in~eq.(\ref{eq:FS}) are the classic  second-order Doppler effect and gravitational red-shift \footnote{One can notice that the
separation between a gravitational red-shift and a Doppler effect is specific to
the chosen coordinate system. One can read the book by Synge~\cite{Synge1960} for a different interpretation in terms of relative velocity  and Doppler effect
only.}. Terms of order $c^{-3}$ amount to less than $3.6\times 10^{-14}$ for a
satellite in Low-Earth Orbit and $2.2\times 10^{-15}$ for the ground.
Terms of order $c^{-4}$ omitted in eq.~(\ref{eq:FS}) can reach a few parts in $10^{-19}$ in the vicinity of the Earth~\cite{Wolf1995}.

\subsubsection{Fibre propagation comparisons}
\label{sec:fibre_prop}
If the signal propagates in an optical fibre, the term~($\dd t_B/\dd t_A$) has
been calculated up to order $c^{-3}$ in~\cite{Gersl2015} for one-way and two-way time and frequency transfers. The result for one-way frequency transfer is:
\begin{equation}
	\frac{\dd t_{B}}{\dd t_A} = 1+\frac{1}{c}\int_0^{L}\left(\frac{\partial n}{\partial t}+n\alpha\frac{\partial T}{\partial t}\right) \dd l + \frac{1}{c^2}\int_0^{L}\frac{\partial \vec{v}\cdot\vec{s}_l}{\partial t} \dd l \ ,
\end{equation}
where $L$ is the total rest length of the fibre at time $t_A$ of emission, $n$ is the effective refractive index of the fibre, $\alpha$ is the linear thermal expansion coefficient of the fibre, $T$ is the temperature of the fibre as a function of time and location, and $\vec{v}$ and $\vec{s}_l$ are the velocity and tangent vector fields of the fibre, respectively.

Up to second order it does not depend on the gravitational field, as for the free propagation in vacuum. The first order term is due to the variation of the fibre length (eg. due to thermal expansion) and of its refractive index. For a 1000~km fibre with
refractive index $n=1.5$ this term is equal to $2\times 10^{-13}$. This term cancels in a two-way frequency transfer. The second
order term is the derivative of the Sagnac effect, which is of order
$10^{-19}$ or less for a 1000~km fibre. The sign of this term depends on the direction of propagation of the signal in the fibre, such that it adds up when doing two-way frequency transfer. Finally the neglected third order term is of the order of $10^{-22}$ for a 1000~km fibre.

\subsection{Clock syntonization}
\label{sec:synto}
Clock syntonization necessitates to calculate the derivatives $(\D\tau/\D t)$ from~(\ref{eq:ratioEM}). Using~(\ref{eq:transfoTT}) and neglecting all terms smaller than $10^{-18}$ we deduce:
\begin{equation}
	\left( \frac{\D \tau}{\D TT} \right)_A^{-1} \left( \frac{\D \tau}{\D TT} \right)_B = 1 + \frac{W_A-W_B}{c^2} \ ,
	\label{eq:synt}
\end{equation}
where $W=W^{\rm{static}} + W^{\rm{temp}}$. Therefore syntonization necessitates the knowledge of the difference of the gravity field between locations $A$ and $B$. Two widely used geodesy techniques can be used to determine the static part of this difference: geometrical levelling and GNSS positions combined with a gravimetric geoid model. Geometrical levelling has the advantage to be very accurate on short distances (typically 0.2--1.0~mm for a 1~km double run levelling) and should be preferred when comparing clocks within the same institute (local comparison). However, geometrical levelling accumulates errors with increasing distance (up to several dm over 1000~km distance) and the GNSS/geoid approach should be preferred for comparisons between different institutes. 

Direct geometrical levelling between two points $A$ and $B$ does not necessitate a point of reference and leads to a high accuracy. However, when determining the height of the clocks w.r.t. the national height system, the reference point of the zero altitude can be very far away from the clocks and therefore the link to the zero altitude may lead to a bias in the determination of the height of the clock. Moreover, the reference point of the zero altitude can be different from one country to the other, because it can be based on different realizations of mean sea level. This leads to the problem of unifying national height systems~\cite{Rulke2013}. The GNSS/geoid approach allows the derivation of the height system bias term for a particular country. It is therefore possible to correct for the bias in the geometrical levelling technique for international clock comparisons. However long distance errors cannot be avoided in geometrical levelling for distant comparisons, for which the GNSS/geoid approach is more adapted. 

The GNSS/geoid method is based on the assumption that the gravitational potential is regular (zero) at infinity. This has the advantage that when using one gravimetric model, the zero origin of the gravitational potential is coherent between all locations covered by the model. High quality regional models exist for Europe and a new one was developed during the course of the ITOC project (EGG2015, see section~\ref{sec:egg}). Indeed, this technique is highly dependent on the quality and coverage of the ground gravimetric observables, and particular care should be taken in the use of the gravimetry dataset. This method allows the derivation of absolute potential values with about 2--3~cm accuracy in terms of heights (best case scenario, i.e., accurate GNSS positions, sufficient terrestrial data around sites of interest, and state-of-the-art global satellite geopotential utilized). Detailed considerations about the uncertainties of the two approaches, geometric levelling and GNSS/geoid, can be found in~\cite{Denker2017}.

\subsection{The realization of Terrestrial Time (TT)}
\label{sec:TT}
The realization of TT necessitates the knowledge of the absolute gravity potential. The atomic international time (TAI) is the most commonly used realization of TT~\cite{Arias2011,Petit2014}. First, comparisons of about 400 atomic clocks around the world in around 70 laboratories are used to calculate the free atomic scale (EAL), a fly-wheel timescale. In a second step, around 15 AFS are used to steer the unit of EAL such that its scale corresponds to the definition of the second. Direct comparisons between AFS are not necessary in this process. Instead, each laboratory compares its AFS to a (master) clock participating in EAL.

In 1980, definition of TAI was given by the Consultative Committee for the Definition of the Second as: 
\begin{quote}
	``TAI is a coordinate time scale defined in a geocentric reference frame with the SI second as realized on the rotating geoid as the scale unit.''
\end{quote}
This reference to the geoid was very ambiguous. Indeed, the value of gravity potential on the geoid, $\Wg$, depends on the global ocean level which changes with time\footnote{Here we use notation $\Wg$ instead of the commonly used $W_0$, in order to emphasize that there is no generally accepted conventional and unified value of the geoid gravity potential value.}. In addition, there are several methods to realize the geoid as ``closest to the mean sea level'' so that there is yet no adopted standard to define a reference geoid and $\Wg$ value (see e.g. a discussion in~\cite{Sanchez2012}). Several authors have considered the time variation of $\Wg$ , see e.g.~\cite{Bursa2006,Dayoub2012}, but there is some uncertainty in what is accounted for in such a linear model. A recent estimate by Dayoub et al. over 1993-2009 gives $\D \Wg / \D t = - 2.7\times10^{-2}$~m$^2$.s$^{-2}$.yr$^{-1}$, mostly driven by the sea level change of +2.9~mm/yr. However, the rate of change of global ocean level could vary during the next decades, and predictions are highly model dependent~\cite{Jevrejeva2012}. Nevertheless, to state an order of magnitude, considering a systematic variation in the sea level of order 2~mm/yr, different definitions of a reference surface for the gravity potential could yield differences in the redshift correction of the AFS of order $2\times10^{-18}$ in a decade, which is of the same order than the best current SFS accuracies~\cite{Huntemann2016,Schioppo2017}.

However, this ambiguity disappeared with the new definition of TT adopted with IAU resolution B1.9 (2000)~\cite{Soffel2003} (see section~\ref{sec:ref_syst}). If TAI is a realization of TT then one has to apply a relative frequency correction, or redshift correction, to the AFS frequency such that 
\begin{equation}
	\frac{\D\tau}{\D TT} = 1 \ ,
	\label{eq:isoTT}
\end{equation}
This equation is exact in GR. Given a model of space-time metric and of the AFS worldlines, it implicitly defines an isochronometric hypersurface, i.e. an hypersurface where all clocks run at the same rate as TT. This hypersurface can be foliated using TT coordinate time as a collection of 3-D hypersurfaces $\Sigma_{TT}$ with constant TT (see section~\ref{sec:observers}). Following eq.~(\ref{eq:transfoTT}), the total correction (to be added) to the AFS relative frequency in order to realize TT is:
\begin{equation}
	\epsilon = - \frac{\WIAU - W}{c^2} \ ,
	\label{eq:corr}
\end{equation}
where $\WIAU=c^2 L_G= \SI[group-separator={,},group-digits=integer]{62636856.000519}{\square\meter\per\square\second}$ and $W$ is the gravity potential at the clock location for the considered epoch. We have seen that it is usual in geodesy to separate the problem of modelization of the gravity field in a static part and a part varying with time (see eq.~(\ref{eq:transfoTT}) and section~\ref{sec:geod}). This splitting is conventional and it should be done with care as several conventions exist (see section~\ref{sec:levelling}). Then the AFS correction $\epsilon$ can be split in a static part and a part varying with time:
\begin{equation}
	\epsilon = \epsilon_{\rm static} + \epsilon_{\rm temp} = - \frac{\WIAU - W^{\rm static}}{c^2} + \frac{W^{\rm temp}}{c^2} \ .
	\label{eq:corr2}
\end{equation}

Keeping only the static part of the gravity field, the problem becomes stationary and the isochronometric hypersurface $\Sigma_{TT}$ is uniquely determined for clocks fixed on the Earth's surface. In the weak gravitational field and low velocity limit, it coincides at the lowest order with the Newtonian equipotential of the gravity field with exact value $\WIAU=c^2 L_G$. Higher order relativistic corrections (terms of order $c^{-4}$ in~$\epsilon$) are of the order $2\times10^{-19}$ or below in the AFS relative frequency~\cite{Muller2007}.

We emphasize that the realization of TT does not necessitate any longer the realization of a geoid. The reference equipotential is just an equipotential with a well defined value, $\WIAU$, which is constant in time and exactly known: its value is a convention. However if this reference equipotential is defined theoretically with no ambiguity, it needs to be realized in the same way as the geoid, leading to inacurracies in its realization, mainly due to the imperfect knowledge of the Earth's mass distribution. In this context, it is interesting to note that clocks in orbit around the Earth are less sensitive to the Earth's gravitational field, and thus to errors in its modelization.

As an illustration, let's take a clock in a satellite following an approximately circular orbit of radius $a$ around the Earth. Approximating the Earth gravitational potential along the satellite trajectory with $V = GM/a$, where $GM$ is the Earth gravitational parameter, then the velocity of the clock is $v=\sqrt{GM/a}$. In order for the clock to realize TT, one needs $V + v^2/2 \approx c^2 L_G$, i.e.~$a\approx\SI{9543}{\kilo\meter}$, and a good knowledge of the trajectory of the satellite. It is shown in~\cite{Wolf1995} that at this altitude the effect of solid Earth tides, ocean tide, polar motion, and changes in the atmospheric pressure are below $10^{-18}$ in fractional frequency. Moreover, tidal effects can be calculated with uncertainties also below $10^{-18}$ in fractional frequency.

The definition of the scalar potential in the context of relativistic reference frames, from which the redshift correction formula~(\ref{eq:corr}) is deduced, is coherent in the Newtonian limit with the assumption done in classical geodesy that the Newtonian potential is regular at infinity. Therefore the GNSS/geoid method is very well adapted to the determination of the redshift corrections in the context of relativistic reference frames. As discussed, when using national height systems one has to calculate corrections such that the assumption of regularity is fulfilled over the area covered by the clock comparisons. This will be illustrated in detail in section~\ref{sec:itoc}.

Finally, according to equation~(\ref{eq:synt}) and~(\ref{eq:corr}), syntonizing two AFS necessitates to determine the relative gravity potential between the locations of both clocks, while the realization of TT necessitates the determination of the absolute gravity potential at the location of the contributing AFS. If the redshift correction~(\ref{eq:corr}) is known for two clocks, it is easy to obtain equation~(\ref{eq:synt}) in order to syntonize them. Therefore, both the problem of syntonization and the realization of TT can be tackled by determining the absolute gravity potential at the locations of the contributing AFS.

\subsection{Temporal variations of the gravity field}
For the temporal variations of the gravity field $W^{\rm{temp}}$, one can refer to~\cite{Voigt2016}, where all corrections bigger than $10^{-18}$ in relative frequency are modelled and evaluated. The dominant effect is the gravity potential variation induced by solid Earth tides, which can be (at most) $10^{-16}$ for clock syntonization on international scales, and $10^{-17}$ for the realization of TT. The second major contributor is the induced signal of ocean tides. However, both solid Earth and ocean tide signals can be modelled down to an accuracy of a few parts in $10^{19}$.

Several other time-variable effects can affect the clock comparisons at the $10^{-18}$ level, such as solid Earth pole tides, non-tidal mass redistributions in the atmosphere, the oceans and the continental water storage, as well as secular signals due to sea level changes and glacial isostatic adjustment. Non-tidal mass redistribution effects on the gravity field are strongly dependent on location and/or weather conditions. As clock comparisons now approach the $10^{-18}$ stability, it will be necessary to develop guidelines in order to include these effects for the syntonization of clocks and their contribution to the realization of TT. Recent analysis of optical clock comparisons have included temporal variations~\cite{Takano2016,Delva2017e}.


\section{Geodetic methods for determining the gravity potential}
\label{sec:geod}

This section describes geodetic methods for determining the gravity potential, needed for the computation of the relativistic redshift corrections for optical clock observations. The focus is on the determination of the static (spatially variable) part of the potential field, while temporal variations in the station coordinates and the potential quantities (with the largest components resulting from solid Earth and ocean tide effects, see \cite{Voigt2016}) are assumed to be taken into account through appropriate reductions or by using sufficiently long averaging times. This is common geodetic practice and leads to a quasi-static state (e.g. by referring all quantities to a given epoch), such that the Earth can be considered as a rigid and non-deformable body, uniformly rotating about a body-fixed axis. Hence, all gravity field quantities including the level surfaces are considered in the following as static quantities, which do not change in time. On this basis, the static and temporal components of the gravity potential can be added to obtain the actual potential value at time $t$, as needed, e.g., for the evaluation of clock comparison experiments. 

In this context, a note on the handling of the permanent (time-independent) parts of the tidal corrections is appropriate; for details, see, e.g., \cite{Makinen2009}, \cite{Ihde2008}, or \cite{Denker2013}. The International Association of Geodesy (IAG) has recommended that the so-called ``zero-tide system'' should be used (resolutions no.~9 and 16 from the year 1983; cf.~\cite{bulletin1984}), where the direct (permanent) tide effects are removed, but the indirect deformation effects associated with the permanent tidal deformation are retained. Unfortunately, geodesy and other disciplines do not strictly follow the IAG resolutions for the handling of the permanent tidal effects, and therefore, depending on the application, appropriate corrections may be necessary to refer all quantities to a common tidal system (see below and the aforementioned references).

In the following, some fundamentals of physical geodesy are given, and then two geodetic methods are described for determining the gravity potential, considering both the geometric levelling approach and the GNSS/geoid approach (GNSS -- Global Navigation Satellite Systems), together with corresponding uncertainty considerations.

\subsection{Fundamentals of physical geodesy}

Classical physical geodesy is largely based on the Newtonian theory with Newton's law of gravitation, giving the gravitational force between two point masses, to which a gravitational acceleration (also termed gravitation) can be ascribed by setting the mass at the attracted point $P$ to unity. Then, by the law of superposition, the gravitational acceleration of an extended body like the Earth can be computed as the vector sum of the accelerations generated by the individual point masses (or mass elements), yielding
\begin{equation}
	\vv{b} = \vv{b} \left( \vv{r} \right) = -G \iiint_{\rm{Earth}} \frac{\vv{r}-\vv{r'}}{\left| \vv{r}-\vv{r'} \right|^3} d m \ , \ d m = \rho d v \ , \ \rho=\rho(\vv{r'}) \ , 
	\label{eq:acc}
\end{equation}
where $\vv{r}$ and $\vv{r'}$ are the position vectors of the attracted point $P$ and the source point $Q$, respectively, $d m$ is the differential mass element, $\rho$ is the volume density, $d v$ is the volume element, and $G$ is the gravitational constant. The SI unit of acceleration is \mss, but the non-SI unit Gal is still frequently used in geodesy and geophysics (1~Gal = \SI{0.01}{\metre\per\square\second}, 1~mGal = \SI{1e-5}{\metre\per\square\second}). While an artificial satellite is only affected by gravitation, a body rotating with the Earth also experiences a centrifugal force and a corresponding centrifugal acceleration $\vv{z}$, which is directed outwards and perpendicular to the rotation axis:
\begin{equation}
	\vv{z} = \vv{z}(\vv{p}) = \omega^2 \vv{p} \ .
\end{equation}
In the above equation, $\omega$ is the angular velocity, and $\vv{p}$ is the distance vector from the rotation axis. Finally, the gravity acceleration (or gravity) vector $\vv{g}$ is the resultant of the gravitation $\vv{b}$ and the centrifugal acceleration $\vv{z}$:
\begin{equation}
	\vv{g} = \vv{b} + \vv{z} \ .
\end{equation}

As the gravitational and centrifugal acceleration vectors $\vv{b}$ and $\vv{z}$ both form conservative vector fields or potential fields, these can be represented as the gradient of corresponding potential functions by
\begin{equation}
	\vv{g} = \vv{\nabla} W = \vv{b} + \vv{z} = \vv{\nabla} V_E + \vv{\nabla} Z_E = \vv{\nabla} (V_E + Z_E) \ ,
	\label{eq:gravity}
\end{equation}
where $W$ is the gravity potential, consisting of the gravitational potential $V_E$ and the centrifugal potential $Z_E$. Based on equations (\ref{eq:acc}) to (\ref{eq:gravity}), the gravity potential $W$ can be expressed as
\begin{equation}
	W = W(\vv{r}) = V_E + Z_E = G \iiint_{\rm{Earth}} \frac{\rho dv}{l} + \frac{\omega^2}{2} p^2 \ ,
	\label{eq:gravity2}
\end{equation}
where $l$ and $p$ are the lengths of the vectors $\vv{r} - \vv{r'}$ and $\vv{p}$, respectively. All potentials are defined with a positive sign, which is common geodetic practice. The gravitational potential $V_E$ is assumed to be regular (i.e. zero) at infinity and has the important property that it fulfills the Laplace equation outside the masses; hence it can be represented by harmonic functions in free space, with the spherical harmonic expansion playing a very important role. Further details on potential theory and properties of the potential functions can be found, e.g., in \cite{Heiskanen1967,Torge2012,Denker2013}.

The determination of the gravity potential $W$ as a function of position is one of the primary goals of physical geodesy; if $W(\vv{r})$ were known, then all other parameters of interest could be derived from it, including the gravity vector $\vv{g}$ according to equation~(\ref{eq:gravity}) as well as the form of the equipotential surfaces (by solving the equation $W(\vv{r}) = \mathrm{const}$.). Furthermore, the gravity potential is also the ideal quantity for describing the direction of water flow, i.e.~water flows from points with lower gravity potential to points with higher values. However, although the above equation  is fundamental in geodesy, it cannot be used directly to compute the gravity potential $W$ due to insufficient knowledge about the density structure of the entire Earth; this is evident from the fact that densities are at best known with two to three significant digits, while geodesy generally strives for a relative uncertainty of at least $10^{-9}$ for all relevant quantities (including the potential $W$). Therefore, the determination of the exterior potential field must be solved indirectly based on measurements performed at or above the Earth's surface, which leads to the area of geodetic boundary value problems (GBVPs; see below).

The gravity potential is closely related to the question of heights as well as level or equipotential surfaces and the geoid, where the geoid is classically defined as a selected level surface with constant gravity potential $W_0$, conceptually chosen to approximate (in a mathematical sense) the mean ocean surface or mean sea level (MSL). However, MSL does not coincide with a level surface due to the forcing of the oceans by winds, atmospheric pressure, and buoyancy in combination with gravity and the Earth's rotation. The deviation of MSL from a best fitting equipotential surface (geoid) is denoted as the (mean) dynamic ocean topography (DOT); it reaches maximum values of about \SI{+-2}{\meter} and is of vital importance for oceanographers for deriving ocean circulation models~\cite{Condi2004}.

On the other hand, a substantially different approach was chosen by the IAG during its General Assembly in Prague, 2015, within ``IAG Resolution (No. 1) for the definition and realization of an International Height Reference System (IHRS)'' \cite{Drewes2016}, where a numerical value $\WIHRS = \SI[group-separator={,}]{62636853.4}{\square\meter\per\square\second}$ (based on observations and data related to the mean tidal system) is defined for the realization of the IHRS vertical reference level surface, with a corresponding note, stating that $\WIHRS$ is related to ``the equipotential surface that coincides (in the least-squares sense) with the worldwide mean ocean surface, the most accepted definition of the geoid'' \cite{Ihde2015}. Although the classical geodetic geoid definition and the IAG 2015 resolution both refer to the worldwide mean ocean surface, so far no adopted standards exist for the definition of MSL, the handling of time-dependent terms (e.g., due to global sea level rise), and the derivation of $W_0$, where the latter value can be determined in principle from satellite altimetry and a global geopotential model (see \cite{Bursa1999,Sanchez2016}). Furthermore, the IHRS value for the reference potential is inconsistent with the corresponding value $\WIAU$ used for the definition of TT (see Sect.~\ref{sec:TT}); Petit et al.~\cite{Petit2014} denote these two definitions as ``classical geoid''  and ``chronometric geoid'', respectively.

In this context, it is somewhat unfortunate that the same notation ($W_0$) is used to represent different estimates for a quantity that is connected with the (time-variable) mean ocean surface, but this issue can be resolved only through future international cooperation, even though it seems unlikely that the different communities are willing to change their definitions. In the meantime, this problem has to be handled by a simple constant shift transformation between the different level surfaces, associated with a thorough documentation of the procedures and conventions involved. It is clear that the definition of the zero level surface ($W_0$ issue) is largely a matter of convention, where a good option is probably to select a conventional value of $W_0$ (referring to a certain epoch) with a corresponding zero level surface, and to describe then the potential of the time-variable mean ocean surface for any given point in time as the deviation from this reference value.

\subsection{The geometric levelling approach}

\begin{figure}[t]
	\centering
	\includegraphics[width=\linewidth]{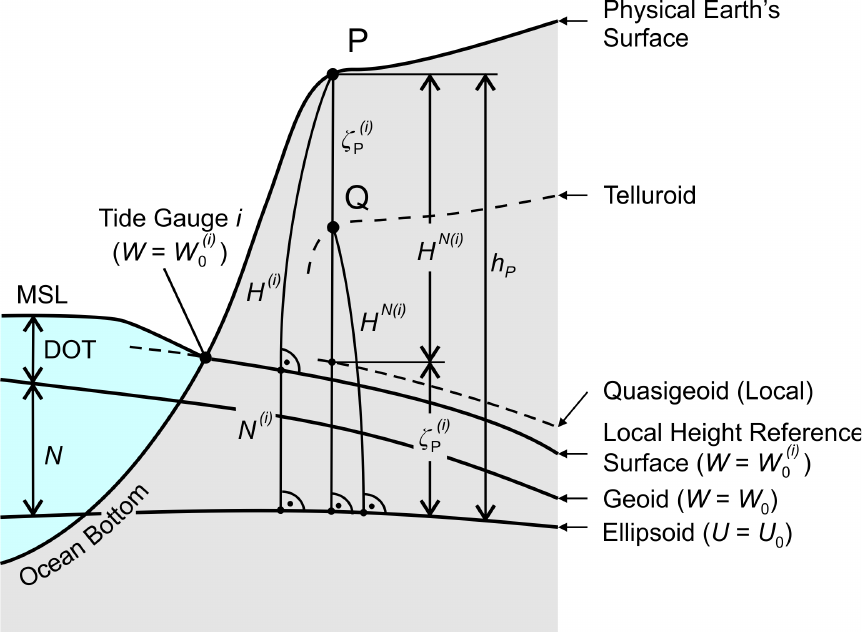}
	\caption{Illustration of several quantities involved in gravity field modelling.}
	\label{fig:illustr}
\end{figure}

The classical and most direct way to obtain gravity potential differences is based on geometric levelling and gravity observations, denoted here as the geometric levelling approach. Based on equation~(\ref{eq:gravity}), the gravity potential differential can be expressed as
\begin{equation}
	d W = \frac{\partial W}{\partial x} d x + \frac{\partial W}{\partial y} d y + \frac{\partial W}{\partial z} d z = \vv{\nabla}W \cdot \vD s = \vv{g} \cdot \vD s = -g \ dn \ ,
\end{equation}
where $\vD s$ is the vectorial line element, $g$ is the magnitude of the gravity vector, and $d n$ is the distance along the outer normal of the level surface (zenith or vertical), which by integration leads to the geopotential number $C$ in the form
\begin{equation}
	\Ci = \Wzi - W_P = - \int_{P_{0(i)}}^{P} dW = \int_{P_{0(i)}}^{P} g \ dn \ ,
	\label{eq:geopot}
\end{equation}
where $P$ is a point at the Earth's surface, $(i)$ refers to the selected zero level or height reference surface (height datum) with the gravity potential $\Wzi$, and $P_{0(i)}$ is an arbitrary point on that level surface. Thus, in addition to the raw levelling results ($d n$), gravity observations ($g$) are needed along the path between $P_{0(i)}$ and $P$, for details, see, e.g.,~\cite{Torge2012}. The geopotential number $C$ is defined such that it is positive for points $P$ above the zero level surface, similar to heights. It should be noted that the integral in eq.~(\ref{eq:geopot}) and hence $C$ is path independent, as the gravity field is conservative. Furthermore, the geopotential numbers can be directly linked to the redshift correction according to eq.~(\ref{eq:corr2}) if one takes $\WIAU$ as zero reference zero level reference potential.

However, regarding height networks, the zero level surface and the corresponding potential is typically selected in an implicit way by connecting the levelling to a fundamental national tide gauge, but the exact numerical value of the reference potential  is usually unknown. As mean sea level deviates from a level surface within the Earth's gravity field due to the dynamic ocean topography (see Fig.~\ref{fig:illustr}), this leads to inconsistencies of more than \SI{0.5}{\meter} between different national height systems across Europe, the extreme being Belgium, which differs by more than \SI{2}{\meter} from all other European countries due to the selection of low tide water as the reference (instead of mean sea level).

Geometric levelling (also called spirit levelling) itself is a quasi-differential technique, which provides height differences $\delta n$ (backsight minus foresight reading) with respect to a local horizontal line of sight. The uncertainty of geometric levelling is rather low over shorter distances, where it can reach the sub-millimetre level, but it is susceptible to systematic errors up to the decimetre level over \SI{1000}{\kilo\meter} distance (see Sect.~\ref{sec:unc}). In addition, the non-parallelism of the level surfaces cannot be neglected over larger distances, as it results in a path dependence of the raw levelling results ($\oint dn \ne 0$), but this problem can be overcome by using potential differences, which are path independent because the gravity field is conservative ($\oint dW = 0$). For this reason, geopotential numbers are almost exclusively used as the foundation for national and continental height reference systems (vertical datum) worldwide, but one can also work with heights and corresponding gravity corrections to the raw levelling results (cf.~\cite{Torge2012}).

Although the geopotential numbers are ideal quantities for describing the direction of water flow, they have the unit \mmss{} and are thus somewhat inconvenient in disciplines such as civil engineering. A conversion to metric heights is therefore desirable, which can be achieved by dividing the $C$ values by an appropriate gravity value. Widely used are the orthometric heights (e.g. in the USA, Canada, Austria, and Switzerland) and normal heights (e.g.~in Germany, France and many other European countries). Heights also play an important role in gravity field modelling due to the strong height dependence of various gravity field quantities.

The orthometric height $H$ is defined as the distance between the surface point $P$ and the zero level surface (geoid), measured along the curved plumb line, which  explains the common understanding of this term as ``height above sea level''~\cite{Torge2012}. All relevant height and gravity field related quantities, are illustrated in figure~\ref{fig:illustr}. The orthometric height can be derived from equation~(\ref{eq:geopot}) by integrating along the plumb line, giving
\begin{equation}
	\Hi=\frac{\Ci}{\bar{g}} \ , \ \bar{g}=\frac{1}{\Hi}\int_0^{\Hi}g \ dH \ ,
	\label{eq:ortho}
\end{equation}
where $\bar{g}$ is the mean gravity along the plumb line (inside the Earth). As $\bar{g}$ cannot be observed directly, hypotheses about the interior gravity field are necessary, which is one of the main drawbacks of the orthometric heights. Therefore, in order to avoid hypotheses about the Earth's interior gravity field, the normal heights $H^{N}$ were introduced by Molodensky (e.g. \cite{Molodensky1962}) in the form
\begin{equation}
	\HNi=\frac{\Ci}{\bar{\gamma}} \ , \ \bar{\gamma}=\frac{1}{\HNi}\int_0^{\HNi}\gamma \ d H^{N} \ ,
	\label{eq:normal}
\end{equation}
where $\bar\gamma$ is a mean normal gravity value along the normal plumb line (within the normal gravity field, associated with the level ellipsoid), and $\gamma$ is the normal gravity acceleration along this line. Consequently, the normal height $H^{N}$ is measured along the slightly curved normal plumb line~\cite{Torge2012}. This definition avoids hypotheses about the Earth's interior gravity field, which is the main reason for adopting it in many countries. Indeed, the value $\bar\gamma$ can be calculated analytically, as the normal gravity potential of the level ellipsoid $U$ is known analytically (see next section), but $\bar\gamma$ is slightly depending on the chosen reference ellipsoid. However, the normal height does not have a simple physical interpretation, in contrast to the orthometric height (``height above sea level''). Nevertheless, the normal height can be interpreted as the height above the quasigeoid, which is not a level surface and also has no physical interpretation (see~\cite{Torge2012}).

While the orthometric and normal heights are related to the Earth's gravity field (so-called physical heights), the ellipsoidal heights $h$, as derived from GNSS observations, are purely geometric quantities, describing the distance (along the ellipsoid normal) of a point $P$ from a conventional reference ellipsoid. As the geoid and quasigeoid serve as the zero height reference surface (vertical datum) for the orthometric and normal heights, respectively, the following relation holds
\begin{equation}
	h = \Hi+\Ni=\HNi+\zi \ ,
	\label{eq:undulation}
\end{equation}
where $\Ni$ is the geoid undulation, and $\zi$ is the quasigeoid height or height anomaly; for further details on the geoid and quasigeoid (height anomalies) see, e.g., \cite{Torge2012}. Equation (\ref{eq:undulation}) neglects the fact that strictly the relevant quantities are measured along slightly different lines in space, but the maximum effect is only at the sub-millimetre level (for further details cf. \cite{Denker2013}). 

Lastly, the geometric levelling approach gives only gravity potential differences, but the associated constant zero potential $\Wzi$ can be determined by at least one (better several) GNSS and levelling points in combination with the (gravimetrically derived) disturbing potential, as described in the next section. Rearranging the above equations gives the desired gravity potential values in the form
\begin{equation}
	W_P = \Wzi - \Ci = \Wzi - \bar{g}\Hi = \Wzi - \bar\gamma \HNi \ ,
	\label{eq:pot2}
\end{equation}
and hence the geopotential numbers and the heights $\Hi$ and $\HNi$ are fully equivalent.

\subsection{The GNSS/geoid approach}
\label{sec:GNSS_geoid}
The gravity potential $W$ cannot be derived directly from equation~(\ref{eq:gravity2}) due to insufficient knowledge about the density structure of the entire Earth, and therefore it must be determined indirectly based on measurements performed at or above the Earth's surface, which leads to the area of geodetic boundary value problems. In this context, gravity measurements form one of the most important data sets. However, since gravity (represented as $g = |\vv{g}| =$ length of the gravity vector $\vv{g}$) and other relevant observations depend in general in a nonlinear way on the potential $W$, the observation equations must be linearized. This is done by introducing an a priori known reference potential and corresponding reference positions. Regarding the reference potential, traditionally the normal gravity field related to the level ellipsoid is employed, where the ellipsoid surface is a level surface of its own gravity field. The level ellipsoid is chosen as a conventional system, because it is easy to compute (from just four fundamental parameters; e.g. two geometrical parameters for the ellipsoid plus the total mass $M$ and the angular velocity $\omega$), useful for other disciplines, and also utilized for describing station positions (e.g. in connection with GNSS or the International Terrestrial Reference Frame -- ITRF). However, today spherical harmonic expansions based on satellite data could also be employed (cf.~\cite{Denker2013}).

The linearization process leads to the disturbing (or anomalous) potential $T$ defined as
\begin{equation}
	T_P = W_P - U_P \ ,
	\label{eq:anom_pot}
\end{equation}
where $U$ is the normal gravity potential associated with the level ellipsoid. Accordingly, the gravity vector and other gravity field observables are approximated by corresponding reference quantities based on the level ellipsoid, leading to gravity anomalies $\Delta g$, height anomalies $ \zeta $, geoid undulations $N$, etc. The main advantage of the linearization process is that the residual quantities (with respect to the known ellipsoidal reference field) are in general four to five orders of magnitude smaller than the original ones, and in addition they are less position dependent.

Hence, the disturbing potential $T$ takes over the role of $W$ as the new fundamental target quantity, to which all other gravity field quantities of interest are related. Accordingly, the gravity anomaly is given by
\begin{equation}
	\Delta g_P = g_P - \gamma_Q = - \frac{\partial T}{\partial h} + \frac{1}{\gamma} \frac{\partial\gamma}{\partial h} T - \frac{\partial\gamma}{\partial h} \left( \Wzi - U_0 \right) \ ,
	\label{eq:grav_anom}
\end{equation}
where $g_P$ is the gravity acceleration at the observation point $P$ (at the Earth's surface or above), $ \gamma_Q$ is the normal gravity acceleration at a known linearization point $Q$ (telluroid, $Q$ is located on the same ellipsoidal normal as $P$ at a distance $H^N$ above the ellipsoid, or equivalently $U_Q = W_P$; for further details, see \cite{Denker2013}), the partial derivatives are with respect to the ellipsoidal height $h$, and $\delta\Wzi = \Wzi - U_0$ is the potential difference between the zero level height reference surface ($\Wzi$) and the normal gravity potential $U_0$ at the surface of the level ellipsoid. Eq.~(\ref{eq:grav_anom}) is also denoted as the fundamental equation of physical geodesy; it represents a boundary condition that has to be fulfilled by solutions of the Laplace equation for the disturbing potential $T$, sought within the framework of GBVPs. Moreover, the subscripts $P$ and $Q$ are dropped on the right side of Eq.~(\ref{eq:grav_anom}), noting that it must be evaluated at the known telluroid point (boundary surface). 

In a similar way, Bruns's formula gives the height anomaly or quasigeoid height as a function of $T$ in the form
\begin{equation}
	\zi = h-\HNi = \frac{T}{\gamma} - \frac{\Wzi-U_0}{\gamma} = \frac{T}{\gamma} - \frac{\delta\Wzi}{\gamma} = \zeta+\zzi \ ,
	\label{eq:anomaly}
\end{equation}
implying that $\zi$ and $\zeta$ are associated with the corresponding zero level surfaces $W = \Wzi$  and $W=U_0$, respectively. The $\delta\Wzi$ term is also denoted as height system bias and is frequently omitted in the literature, implicitly assuming that $\Wzi$ equals $U_0$. However, when aiming at a consistent derivation of absolute potential values, the $\delta\Wzi$ term has to be taken into consideration.

Hence, all linearized gravity field observables are linked to the disturbing potential~$T$, which has the important property of being harmonic outside the Earth's surface and regular (zero) at infinity. Consequently, solutions of $T$ are developed in the framework of potential theory and GBVPs, i.e. solutions of the Laplace equation are sought that fulfil certain boundary conditions. Now, the first option to compute $T$ is based on the well-known spherical harmonic expansion, using coefficients derived from satellite data alone or in combination with terrestrial data (e.g., EGM2008; EGM -- Earth Gravitational Model~\cite{Pavlis2012}), yielding
\begin{equation}
	T(\theta,\lambda,r) = \sum_{n=0}^{n_{ \mathrm{max} }} \left( \frac{a}{r} \right)^{n+1} \sum_{m=-n}^{n} \overline T_{nm} \overline Y_{nm} (\theta,\lambda)
	\label{eq:EGM}
\end{equation}
with
\begin{eqnarray}
	\overline Y_{nm}(\theta,\lambda) = \overline P_{n|m|}(\cos\theta) \left\{
		\begin{array}{l}
        		\cos m\lambda \ \mathrm{for} \ m\ge0\\
			\sin |m|\lambda \ \mathrm{for} \ m<0
		\end{array}
	\right. \ , \label{eq:Ynm} \\
	\overline T_{nm} = \frac{GM}{a} \left\{
		\begin{array}{l}
			\Delta \overline C_{nm} \ \mathrm{for} \ m\ge0 \\
			\Delta \overline S_{nm} \ \mathrm{for} \ m<0
		\end{array}
	\right. \ ,
	\label{eq:Tnm}
\end{eqnarray}
where $(\theta,\lambda,r)$ are spherical coordinates, $n$, $m$ are integers denoting the degree and order, $GM$ is the geocentric gravitational constant (gravitational constant $G$ times the mass of the Earth $M$), $a$ is in the first instance an arbitrary constant, but is typically set equal to the semimajor axis of a reference ellipsoid, $\overline P_{nm}(\cos\theta)$ are the fully normalized associated Legendre functions of the first kind, and $\Delta \overline C_{nm}$, $\Delta \overline S_{nm}$ are the (fully normalized) spherical harmonic coefficients (also denoted as Stokes's constants), representing the difference in the gravitational potential between the real Earth and the level ellipsoid.

Regarding the uncertainty of a gravity field quantity computed from a global spherical harmonic model up to some fixed degree $n_{\mathrm{max}}$, the coefficient uncertainties lead to the so-called commission error based on the law of error propagation, and the omitted coefficients above degree $n_{\mathrm{max}}$, which are not available in the model, lead to the corresponding omission error. With dedicated satellite gravity field missions such as GRACE (Gravity Recovery and Climate Experiment) and GOCE (Gravity field and steady-state Ocean Circulation Explorer), the long wavelength geoid and quasigeoid can today be determined with low uncertainty, e.g., about \SI{1}{\milli\meter} at \SI{200}{\kilo\meter} resolution ($n= 95$) and \SI{1}{\centi\meter} at \SI{150}{\kilo\meter} resolution (\textit{n }= 135) from GRACE (e.g.~\cite{Mayer-Gurr2014}), and \SI{1.5}{\centi\meter} at about \SI{110}{\kilo\meter} resolution ($n = 185$) from the GOCE mission (e.g. \cite{Mayer2015,Brockmann2014}). However the corresponding omission error at these wavelengths is still quite significant with values at the level of several decimetres, e.g., \SI{0.94}{\meter} for $n = 90$, \SI{0.42}{\meter} for $n = 200$, and \SI{0.23}{\meter} for $n = 360$. For the ultra-high degree geopotential model EGM2008~\cite{Pavlis2012}, which combines satellite and terrestrial data and is complete up to degree and order 2159, the omission error is \SI{0.023}{\meter}, while the commission error is about~5 to \SI{20}{\centi\meter}, depending on the region and the corresponding data quality. The above uncertainty estimates are based on the published potential coefficient standard deviations as well as a statistical model for the estimation of corresponding omission errors, but do not include the uncertainty contribution of $GM$ (zero degree term in Eq.~(\ref{eq:EGM})); hence, the latter term, contributing about \SI{3}{\milli\meter} in terms of the height anomaly (corresponds  to about 0.5~ppb; see \cite{Smith2000,Ries2014}), has to be added in quadrature to the figures given above. Further details on the uncertainty estimates can be found in~\cite{Denker2013}.

Based on these considerations it is clear that satellite measurements alone will never be able to supply the complete geopotential field with sufficient accuracy, which is due to the signal attenuation with height and the required satellite altitudes of a few \SI{100}{\kilo\meter}. Only a combination of the highly accurate and homogeneous (long wavelength) satellite gravity fields with high-resolution terrestrial data (mainly gravity and topography data with a resolution down to 1--2~km and below) can cope with this task. In this respect, the satellite and terrestrial data complement each other in an ideal way, with the satellite data accurately providing the long wavelength field structures, while the terrestrial data sets, which have potential weaknesses in large-scale accuracy and coverage, mainly contribute the short wavelength features. However, in the future, also height anomalies derived from common GNSS and clock points may contribute to regional gravity field modelling (see Sect.~\ref{sec:high_res}).

Consequently, regional solutions for the disturbing potential and other gravity field parameters have to be developed, which typically have a higher resolution (down to 1--2 km) than global spherical harmonic models. Based on the developments of Molodensky (e.g., \cite{Molodensky1962}), the disturbing potential $T$ can be derived from a series of surface integrals, involving gravity anomalies and heights over the entire Earth's surface, which in the first instance can be symbolically written as
\begin{equation}
	T=\mathbf{M}(\Delta g) \ ,
	\label{eq:molo}
\end{equation}
where $\mathbf{M}$ is the Molodensky operator and $\Delta g$ are the gravity anomalies over the entire Earth's surface.

Further details on regional gravity field modelling are given in~\cite{Denker2013,Denker2017}, including the solution of Molodensky's problem, the remove-compute-restore (RCR) procedure, the spectral combination approach, data requirements, and uncertainty estimates for the disturbing potential and quasigeoid heights. These investigations show that quasigeoid heights can be obtained today with an estimated uncertainty of \SI{1.9}{\centi\meter}, where the major contributions come from the spectral band below spherical harmonic degree 360. Furthermore, this uncertainty estimate represents an optimistic scenario and is only valid for the case that a state-of-the-art global satellite model (e.g.~a 5th generation GOCE model~\cite{Brockmann2014}) is employed and sufficient high-resolution and high-quality terrestrial gravity and terrain data sets (especially gravity measurements with a spacing of a few kilometers and an uncertainty lower than 1~mGal) are available around the point of interest (e.g.~within a distance of 50--100~km), see also~\cite{Forsberg1993,Jekeli2012}. Fortunately, such a data situation exists for most of the metrology institutes with optical clock laboratories -- at least in Europe. Furthermore, the perspective exists to improve the uncertainty of the calculated quasigeoid heights~\cite{Denker2017}.

Now, once the disturbing potential values $T$ are computed, either from a global geopotential model by equation~(\ref{eq:EGM}), or from a regional solution by equation~(\ref{eq:molo}) based on Molodensky's theory, the gravity potential $W$, needed for the relativistic redshift corrections, can be computed most straightforwardly from eq.~(\ref{eq:anom_pot}) as
\begin{equation}
	W_P = U_P + T_P \ ,
	\label{eq:gravP}
\end{equation}
where the basic requirement is that the position of the given point $P$ in space must be known accurately (e.g. from GNSS observations), as the normal potential $U$ is strongly height-dependent, while $T$ is only weakly height dependent with a maximum vertical gradient of a few parts in $10^{-3}$~\mmss{} per metre. The above equation also makes clear that the predicted potential values $W_P$ are in the end independent of the choice of $W_0$ and $U_0$ used for the linearization. Furthermore, by combining equation~(\ref{eq:gravP}) with~(\ref{eq:anomaly}), and representing $U$ as a function of $U_0$ and the ellipsoidal height~$h$, the following alternative expressions for $W$ (at point $P$) can be derived as
\begin{equation}
	W_P = U_0 - \bar{\gamma} (h-\zeta) = U_0 - \bar{\gamma} \left( h-\zi \right) + \delta\Wzi \ ,
	\label{eq:pot1}
\end{equation}
which demonstrates that ellipsoidal heights (e.g. from GNSS) and the results from gravity field modelling in the form of the quasigeoid heights (height anomalies) $\zeta$ or the disturbing potential $T$ are required, whereby a similar equation can be derived for the geoid undulations $N$. Consequently, the above approach (equations~(\ref{eq:gravP}) and~(\ref{eq:pot1})) is denoted here somewhat loosely as the GNSS/geoid approach, which is also known in the literature as the GNSS/GBVP approach (the geodetic boundary value problem is the basis for computing the disturbing potential $T$; see, e.g., \cite{Rummel1988,Heck1990}). 

The GNSS/geoid approach depends strongly on precise gravity field modelling (disturbing potential $T$, metric height anomalies $\zeta$ or geoid undulations $N$) and precise GNSS positions (ellipsoidal heights $h$) for the points of interest, with the advantage that it delivers the absolute gravity potential $W$, which is not directly observable and is therefore always based on the assumption that the gravitational potential is regular (zero) at infinity (see above). In addition, the GNSS/geoid approach allows the derivation of the height system bias term $\delta\Wzi$ based on equation (\ref{eq:anomaly}) together with at least one (better several) common GNSS and levelling stations in combination with the gravimetrically determined disturbing potential $T$.

\subsection{Uncertainty considerations}
\label{sec:unc}

The following uncertainty considerations are based on heights, but corresponding potential values can easily be obtained by multiplying the meter values with an average gravity value (e.g.~\SI{9.81}{\meter\per\square\second} or roughly~\SI{10}{\meter\per\square\second}). Regarding the geometric levelling and the GNSS/geoid approach, the most direct and accurate way to derive potential differences over short distances is the geometric levelling technique, as standard deviations of 0.2--\SI{1.0}{\milli\meter} can be attained for a \SI{1}{\kilo\meter} double-run levelling with appropriate technical equipment~\cite{Torge2012}. However, the uncertainty of geometric levelling depends on many factors, with some of the levelling errors behaving in a random manner and propagating with the square root of the number of individual set-ups or the distance, respectively, while other errors of systematic type may propagate with distance in a less favourable way. Consequently, it is important to keep in mind that geometric levelling is a differential technique and hence may be susceptible to systematic errors; examples include the differences between the second and third geodetic levelling in Great Britain (about \SI{0.2}{\meter} in the north--south direction over about \SI{1000}{\kilo\meter} distance~\cite{Kelsey1972}), corresponding differences between an old and new levelling in France (about \SI{0.25}{\meter} from the Mediterranean Sea to the North Sea, also mainly in north--south direction, distance about \SI{900}{\kilo\meter}~\cite{Rebischung2008}), and inconsistencies of more than \SI{+-1}{\meter} across Canada and the USA (differences between different levellings and with respect to an accurate geoid~\cite{Veronneau2006,Smith2010,Smith2013}). In addition, a further complication with geometric levelling in different countries is that the results are usually based on different tide gauges with offsets between the corresponding zero level surfaces, which, for example, reach more than 0.5~m across Europe. Furthermore, in some countries the levelling observations are about 100~years old and thus may not represent the actual situation due to possibly occurring recent vertical crustal movements.

With respect to the GNSS/geoid approach, the uncertainty of the GNSS positions is today more or less independent of the interstation distance. For instance, the station coordinates provided by the International GNSS Service (IGS) or the IERS (e.g.~ITRF2008) reach vertical accuracies of about 5--\SI{10}{\milli\meter} (cf.~\cite{Altamimi2011,Altamimi2016,Seitz2013}). The uncertainty of the quasigeoid heights (height anomalies) is discussed in the previous subsection, showing that a standard deviation of \SI{1.9}{\centi\meter} is possible in a best-case scenario. Moreover, the values are nearly uncorrelated over longer distances, with a correlation of less than 10~\%  beyond a distance of about \SI{180}{\kilo\meter}~\cite{Denker2017}. Aiming at the determination of the absolute gravity potential $W$ according to equations~(\ref{eq:gravP}) or~(\ref{eq:pot1}), which is the main advantage of the GNSS/geoid technique over the geometric levelling approach, both the uncertainties of GNSS and the quasigeoid have to be considered. Assuming a standard deviation of \SI{1.9}{\centi\meter} for the quasigeoid heights and \SI{1}{\centi\meter} for the GNSS ellipsoidal heights without correlations between both quantities, a standard deviation of \SI{2.2}{\centi\meter} is finally obtained (in terms of heights) for the absolute potential values based on the GNSS/geoid approach. Thus, for contributions of optical clocks to international timescales, which require the absolute potential $W_P$ relative to a conventional zero potential $W_0$ (see Sect.~\ref{sec:TT}), the relativistic redshift correction can be computed with an uncertainty of about \num{2e-18}. This is the case more or less everywhere in the world where high-resolution regional gravity field models have been developed on the basis of a state-of-the-art global satellite model in combination with sufficient terrestrial gravity field data. On the other hand, for potential differences over larger distances of a few \SI{100}{\kilo\meter} (i.e.~typical distances between different metrology institutes), the statistical correlations of the quasigeoid values virtually vanish, which then leads to a standard deviation for the potential difference of 3.2~cm in terms of height, i.e. $\sqrt{2}$ times the figure given above for the absolute potential (according to the law of error propagation), which again has to be considered as a best-case scenario. This would also hold for intercontinental connections between metrology institutes, provided again that sufficient regional high-resolution terrestrial data exist around these places. Furthermore, in view of future refined satellite and terrestrial data, the perspective exists to improve the uncertainty of the relativistic redshift corrections from the level of a few parts in $10^{18}$ to one part in $10^{18}$ or below. According to this, over long distances across national borders, the GNSS/geoid approach should be a better approach than geometric levelling (see also~\cite{Denker2017}).


\section{Relativistic redshift corrections for the realization of TT from geodetic methods}
\label{sec:itoc}

An atomic clock, in order to contribute to the realization of Terrestrial Time (TT), needs to be corrected for the relativistic redshift (see Sect.~\ref{sec:TT}, eq.~(\ref{eq:corr})). In this section we present some results from the ITOC (International Timescales with Optical Clocks) project, in particular those linked to the determination of unified relativistic redshift corrections for several European metrology institutes.

\subsection{The ITOC project}
The ITOC project (\cite{Margolis2013}; see also \url{http://projects.npl.co.uk/itoc/}) was a 3 years~(2013--2016) EURAMET joint research project funded by the European Community's Seventh Framework Programme, ERA-NET Plus. This project was done in the context of a future optical redefinition of the SI second~(see e.g.~\cite{Margolis2014,Riehle2015,Gill2016,Riehle2017}). An extensive programme of comparisons between high accuracy European optical atomic clocks has been performed, verifying the estimated uncertainty budgets of the optical clocks. Relativistic effects influencing clock comparisons have been evaluated at an improved level of accuracy, and the potential benefits that optical clocks could bring to the field of geodesy have been demonstrated.

Several optical frequency ratio measurements as well as independent absolute frequency measurements of optical lattice clocks have been made locally at the following NMIs (National Metrology Institutes): INRIM (Istituto Nazionale di Ricerca Metrologica, Torino, Italy), LNE-SYRTE (Laboratoire national de m\'etrologie et d'essais -- Syst\`eme de R\'ef\'erences Temps-Espace, Paris, France), NPL (National Physical Laboratory, Teddington, UK), and PTB (Physikalisch-technische Bundesanstalt, Braunschweig, Germany), all of whom operate one or more than one type of optical clock, as well as Cesium primary frequency standards (see e.g.~\cite{Falke2014,Tyumenev2016,Lodewyck2016,Margolis2015a}). Distant comparisons have also been performed between the same laboratories with a broadband version of two-way satellite time and frequency transfer (TWSTFT).

A proof-of-principle experiment has been realized to show that the relativistic redshift of optical clocks can be exploited to measure gravity potential differences over medium--long baselines. A transportable $^{87}$Sr optical lattice clock has been developed at PTB~\cite{Koller2017}. It has been transported to the Laboratoire Souterrain de Modane (LSM) in the Fr\'ejus road tunnel through the Alps between France and Italy. There it was compared, using a transportable frequency comb from NPL, to the caesium fountain primary frequency standard at INRIM, via a coherent fibre link and a second optical frequency comb operated by INRIM. A physical model has been formulated to describe the relativistic effects relevant to time and frequency transfer over optical fibre links, and has been used to evaluate the relativistic corrections for the fibre links now in place between NPL, LNE-SYRTE and PTB, as well as to provide guidelines on the importance of exact fibre routing for time and frequency transfer via optical fibre links~(see~\cite{Gersl2015} and section~\ref{sec:fibre_prop}).

Within the ITOC project, the gravity potential has been determined with significantly improved accuracy at the sites participating in optical clock comparisons within the project (INRIM, LNE-SYRTE, LSM, NPL and PTB) by IfE/LUH (Institut f\"ur Erdmessung, Leibniz Universit\"at Hannover). Levelling measurements and gravity surveys have been performed at INRIM, LSM, OBSPARIS, NPL and PTB, the latter including at least one absolute gravity observation on each site. These measurements have been integrated into the existing European gravity database and used for the computation of a new version of the European Gravimetric (Quasi) Geoid, EGG2015~(see~\cite{Denker2015} and Sect.~\ref{sec:egg}). Time-variable gravity potential signals induced by tides and non-tidal mass redistributions have also been calculated for the optical clock comparison sites~\cite{Voigt2016}. Finally, the potential contributions of combined GNSS and optical clock measurements for determining the gravity potential at high spatial resolution have been studied theoretically, which will be presented in section~\ref{sec:high_res}.

\subsection{The GNSS and levelling campaigns}
\label{sec:levelling}

Within the ITOC project, GNSS and levelling observations were performed at the NMIs INRIM, LNE-SYRTE, NPL, and PTB, as well as the collaborator LSM (not an NMI) to calculate the relativistic redshift corrections. First of all, some general recommendations were developed for carrying out the measurements to ensure accuracies in the millimetre range for the levelling results and better than one centimetre for the GNSS (ellipsoidal) heights (see \cite{Denker2017}). In general, it is recommended to install fixed markers in all local laboratories close to the clock tables to allow an easy height transfer to the clocks (e.g.~with a simple spirit level used for building construction), and to connect these markers by geometric levelling with millimetre uncertainty to the existing national levelling networks and at least two (better several) GNSS stations. This is to support local clock comparisons at the highest level, and to apply the GNSS/geoid approach to obtain also the absolute potential values for remote clock comparisons and contributions to international timescales, while at the same time improving the redundancy and allowing a mutual control of GNSS, levelling, and (quasi)geoid data. 

\begin{figure}[t]
	\centering
	\includegraphics{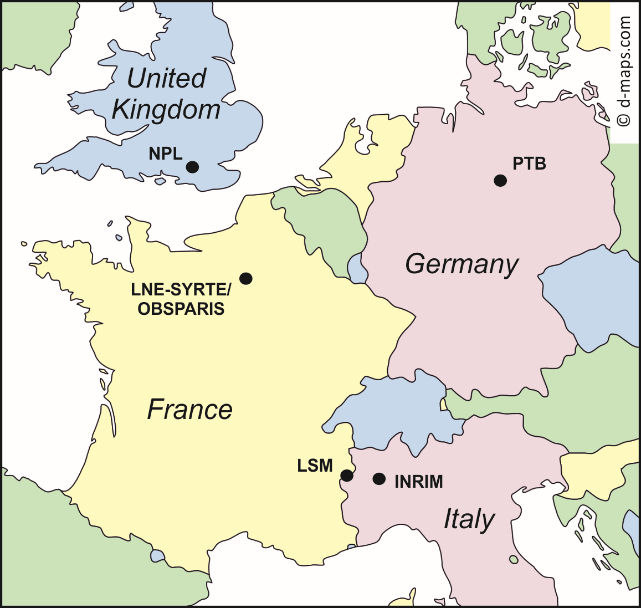}
	\caption{Map showing the locations of the INRIM, LNE-SYRTE, NPL, PTB, and LSM sites}
	\label{fig:map}
\end{figure}

The actual levelling and GNSS measurements were mainly taken by local surveyors on behalf of the respective NMIs, and the NMIs provided all results to Leibniz Universit\"at Hannover (LUH; Institut f\"ur Erdmessung) for further processing and homogenisation. The locations of the above mentioned ITOC clock sites are shown in Fig.~\ref{fig:map}.

The coordinates of all GNSS stations were referred to the ITRF2008 at its associated standard reference epoch 2005.0, with three-dimensional Cartesian and ellipsoidal coordinates being available.~ The geometric levelling results were based in the first instance on the corresponding national vertical reference networks, which are the following:
\begin{itemize}
	\item DHHN92 is the official German height reference system; it is based on the Amsterdam tide gauge and consists of normal heights.
	\item NGF-IGN69 is the official French height reference system; it is based on the Marseille tide gauge and also consists of normal heights. In addition, selected levelling lines were re-observed since 2000, which lead to the so-called (NGF)--NIREF network, differing from the old network mainly by a south-to-north trend of \SI{31.0}{\milli\meter} per degree latitude~\cite{Rebischung2008}.
	\item ODN (Ordnance Datum Newlyn, established by Ordnance Survey) is the height reference system for mainland Great Britain; it is based on the Newlyn tide gauge and consists of orthometric heights.
	\item IGM is the Italian height reference system (established by Istituto Geografico Militare); it is based on the Genova tide gauge and consists of orthometric heights.
\end{itemize}
The different zero level surfaces of the above national height systems (datum) were taken into account by transforming all national heights into the unified European Vertical Reference System (EVRS) using its latest realization EVRF2007 (European Vertical Reference Frame 2007). The EVRF2007 is based on a common adjustment of all available European levelling networks in terms of geopotential numbers, which are finally transformed into normal heights. The measurements within the UELN~(United European Levelling Network) originate from very different epochs, but reductions for vertical crustal movements were only applied for the (still ongoing) post-glacial isostatic adjustment (GIA) in northern Europe; for further details on EVRF2007, see~\cite{Sacher2008}. However, as GIA hardly affects the aforementioned clock sites, while other sources of vertical crustal movements are not known, the EVRF2007 heights are considered as stable in time in the following.

For the conversion of the national heights into the vertical reference frame EVRF2007, nearby common points with heights in both systems were utilized; this information was kindly provided by Bundesamt f\"ur Kartographie und Geod\"asie (BKG) in Germany (M. Sacher, personal communication, 9 October 2015). If such information is not available, the CRS-EU webpage (Coordinate Reference Systems in Europe; \url{http://www.crs-geo.eu}, also operated by BKG) can be used, which gives, besides a description of all the national and international coordinate and height reference systems for the participating European countries, up to three transformation parameters (height bias and two tilt parameters) for the transformation of the national heights into EVRF2007 and a statement on the quality of this transformation.

The EVRF2007 heights were computed as
\begin{equation}
	\HNevrf = H^{\rm (national)} + \Delta H^{\rm (national)} \ ,
	\label{eq:HEVRS}
\end{equation}
where $\Delta H^{\rm (national)}$ is a constant shift for each NMI site. The following offsets $\Delta H^{\rm (national)}$ were employed:\vspace{0.5em}\\
%
\begin{tabular}[h]{lll}
	$\bullet$~PTB:		& \SI{+0.006}{\meter} & (DHHN92),\\
	$\bullet$~LNE-SYRTE:	& \SI{-0.479}{\meter} & (NGF-IGN69),\\
	$\bullet$~NPL:		& \SI{-0.010}{\meter} & (ODN, based on the official EVRF2007 results),\\
			& \SI{-0.144}{\meter} & (ODN, own estimate, see below)\\
	$\bullet$~INRIM:		& \SI{-0.307}{\meter} & (IGM),\\
	$\bullet$~LSM:		& \SI{-0.307}{\meter} & (IGM).\\
\end{tabular}
\vspace{0.5em}\\
The accuracy of the above transformation depends on the accuracy of the input heights as well as the number of identical points, giving RMS residuals of the transformation between \SI{2}{\milli\meter}~(Germany) and \SI{35}{\milli\meter}~(Italy). A further note is necessary for the computation of the NPL offset. The offset of \SI{-0.010}{\meter} is based on the official EVRF2007 heights, which rely on hydrodynamic levelling (see~\cite{Cartwright1963}, but do not include the 1994 channel tunnel levelling. Therefore, a first attempt was made to consider the new channel tunnel levelling as well as the new levelling measurements in France (NIREF, see above); this was done by starting with an offset of \SI{-0.479}{\meter} for LNE-SYRTE, plus a correction for the NGF-IGN69 tilt between LNE-SYRTE and Coquelle (channel tunnel entrance in France) of \SI{-0.065}{\meter} (south-north slope = \SI{-31.0}{\milli\meter} per degree latitude, latitude difference = 3.105~degrees~\cite{Rebischung2008}), plus an offset of \SI{+0.400}{\meter} for the difference between ODN and NGF-IGN69 from the channel tunnel levelling~\cite{Greaves2007}, resulting in an offset of \SI{-0.144}{\meter} for NPL. In Sect.~\ref{sec:grav_pot} it is shown that the new offset leads to a better agreement between the geometric levelling and the GNSS/geoid approach.

Further details on the local levelling results and corresponding GNSS observations at some of the aforementioned clock sites can be found in~\cite{Denker2017}. In general, the uncertainty of the local levellings is at the few millimeters level, and the uncertainty of the GNSS ellipsoidal heights is estimated to be better than \SI{10}{\milli\meter}. Moreover, care has to be taken in the handling of the permanent parts of the tidal corrections (for details, see, e.g., \cite{Makinen2009,Ihde2008,Denker2013}. The IAG has recommended to use the so-called ``zero tide system'' (resolutions no. 9 and 16 from the year 1983; cf.\cite{bulletin1984}), which is implemented in the European height reference frame EVRF2007 and the European gravity field modelling performed at LUH (e.g.~EGG2015, see below). On the other hand, most GNSS coordinates (including the ITRF and IGS results) refer to the ``non-tidal (or tide-free) system''. Hence, for consistency with the IAG recommendations and the other quantities involved (EVRF2007 heights, quasigeoid), the ellipsoidal heights from GNSS were converted from the non-tidal to the zero-tide system based on the following formula from \cite{Ihde2008} with
\begin{equation}
	h_{zt} = h_{nt} + 60.34 - 179.01 \sin^2 \phi - 1.82 \sin^4 \phi \ \ {\rm [mm]} \ ,
	\label{eq:tidal}
\end{equation}
where $\phi$ is the ellipsoidal latitude, and $h_{nt}$ and $h_{zt}$ are the non-tidal and zero-tide ellipsoidal heights, respectively. Hence, the zero-tide heights over Europe are about 3--5~cm smaller than the corresponding non-tidal heights.

\subsection{The European gravimetric quasigeoid model EGG2015}
\label{sec:egg}
The latest European gravimetric quasigeoid model EGG2015~\cite{Denker2015} was employed to determine absolute potential values based on equations~(\ref{eq:gravP}) and~(\ref{eq:pot1}), as needed for the derivation of the relativistic redshift corrections in the context of international timescales. The major differences between EGG2015 and the previous EGG2008 model~\cite{Denker2013} are the inclusion of additional gravity measurements carried out recently around the aforementioned ITOC clock sites~\cite{Margolis2013} and the use of a newer geopotential model based on the GOCE satellite mission instead of EGM2008. The new gravity measurements around the clock sites were carried out by LUH, taking at least one absolute gravity observation (with the LUH FG5X-220 instrument) plus additional relative gravity observations (relative to the established absolute points) around all ITOC sites. The total number of new gravity points is 36 for INRIM, 100 for LNE-SYRTE, 123 for LSM, 66 for NPL, and 46 for PTB, where most of the measurements were taken around LSM due to the high mountains and corresponding strong gravity field variations. Overall, the purpose of the new gravity measurements was threefold, namely to perform spot checks of the (largely historic) gravity data base (consistency check), to add new observations in areas void of gravity data so far (coverage improvement), and to serve for future geodynamic and metrological purposes (infrastructure improvement), with the ultimate goal of improving the reliability and accuracy of the computed quasigeoid model.

EGG2015 was computed from surface gravity data in combination with topographic information and the geopotential model GOCO05S~\cite{Mayer2015} based on the RCR technique. The estimated uncertainty (standard deviation) of the absolute quasigeoid values is \SI{1.9}{\centi\meter}; further details including correlation information can be found in~\cite{Denker2017,Denker2013}. 

\subsection{Gravity potential determination}
\label{sec:grav_pot}

First, a consistency check between the GNSS and levelling heights at each clock site was performed by evaluating the differences between the GNSS ellipsoidal heights and the normal heights from levelling, computed as $\zGNSS = h_{zt} - \HNevrf$, also denoted as GNSS/levelling quasigeoid heights ($h_{zt}$ is referring to ITRF2008, epoch 2005.0, zero-tide system; $\HNevrf$ is based on EVRF2007; see Sect.~\ref{sec:levelling}). As the distances between the GNSS stations at each NMI site are typically only a few \SI{100}{\meter}, the quasigeoid at each site can be approximated in the first instance by a horizontal plane, but a more general and better way (especially for larger interstation distances) is the comparison with a high-resolution gravimetric quasigeoid model, such as EGG2015. After computing the differences $\left( \zGNSS-\zEGG \right)$, the main quantities of interest are the residuals about the mean difference, giving RMS values of \SI{11}{\milli\meter} (max.~\SI{16}{\milli\meter}) for INRIM, \SI{5}{\milli\meter} (max.~\SI{5}{\milli\meter}) for LNE-SYRTE, \SI{6}{\milli\meter} (max.~\SI{8}{\milli\meter}) for NPL, and \SI{4}{\milli\meter} (max.~\SI{6}{\milli\meter}) for PTB, while for LSM only a control through two RTK positions exists, giving a RMS difference of \SI{17}{\milli\meter} (max.~\SI{29}{\milli\meter}); this proves an excellent consistency of the GNSS and levelling results at all clock sites. Although initial results were worse for the PTB and LNE-SYRTE sites, the problem was traced to an incorrect identification of the corresponding antenna reference points (ARPs); at the PTB site, an error of \SI{16}{\milli\meter} was found for station PTBB, and at LNE-SYRTE, there was a difference of \SI{29}{\milli\meter} between the ARP and the levelling benchmark and an additional error in the ARP height of~\SI{8}{\milli\meter} at station OPMT. It should be noted that, due to the high consistency of the GNSS and levelling data at all sites, even quite small problems in the ARP heights (below~\SI{1}{\centi\meter}) could be detected and corrected after on-site inspections and additional verification measurements. This also strongly supports the recommendation to have sufficient redundancy in the GNSS and levelling stations. 

Now, in order to apply the GNSS/geoid approach according to equations~(\ref{eq:gravP}) and~(\ref{eq:pot1}), ellipsoidal heights are required for all stations of interest. However, initially  GNSS coordinates are only available for a few selected points at each NMI site, while for most of the other laboratory points near the clocks, only levelled heights exist. Therefore, based on equation~(\ref{eq:anomaly}), a quantity $\delta\zeta$ is defined as
\begin{equation}
	\delta\zeta = \left( h - \HNi \right) - \zi \ ,
	\label{eq:error}
\end{equation}
which should be zero in theory, but is not in practice due to the uncertainties in the quantities involved (GNSS, levelling, quasigeoid). However, if a high-resolution quasigeoid model is employed (such as EGG2015), the term $\dz$ should be small and represent only long-wavelength features, mainly due to systematic levelling errors over large distances as well as long-wavelength quasigeoid errors. In this case an average (constant) value $\dzb$ (based on the common GNSS and levelling benchmarks) can be used at each NMI site to convert all levelled heights into ellipsoidal heights by using
\begin{equation}
	\hadj = \HNi + \zi + \dzb = \HNi + \left( \zeta+\zzi \right) + \dzb \ ,
	\label{eq:hadj}
\end{equation}
which is based on Eq.~(\ref{eq:undulation}). This has the advantage that locally (at each NMI) the consistency is kept between the levelling results, on the one hand, and the GNSS/quasigeoid results on the other hand. Consequently, the final potential differences between stations at each NMI are identical for the GNSS/geoid and geometric levelling approach, which is reasonable, as locally the uncertainty of levelling is usually lower than that of the GNSS/quasigeoid results. 

Based on the ellipsoidal heights (according to eq.~(\ref{eq:hadj})) and the EVRF2007 normal heights (based on eq.~(\ref{eq:HEVRS})), the gravity potential values can finally be derived for all relevant stations, using both the geometric levelling approach (eq.~(\ref{eq:pot2})) and the GNSS/geoid approach (eqs.~(\ref{eq:gravP}) or~(\ref{eq:pot1})). The results from both approaches are provided in the form of geopotential numbers according to equation~(\ref{eq:geopot}) with
\begin{equation}
	\CIAU = \WIAU - W_P
	\label{eq:ciau}
\end{equation}
where the conventional value $\WIAU=c^2 L_G \approx \SI[group-separator={,},group-digits=integer]{62636856.00}{\square\meter\per\square\second}$ is used, following the IERS2010 conventions and the IAU resolutions for the definition of TT~(see Sect.~\ref{sec:TT}). The geopotential numbers $C$ are more convenient than the absolute potential values \textit{W\textsubscript{P}} due to their smaller numerical values and direct usability for the derivation of the (static) relativistic redshift corrections according to equation~(\ref{eq:corr2}). The geopotential numbers $C$ derived from equation~(\ref{eq:ciau})  are typically given in the geopotential unit (gpu; 1~gpu = \SI{10}{\square\meter\per\square\second}), resulting in numerical values of $C$ that are about 2~\%  smaller than the numerical height values. Regarding the geometric levelling approach, the value $\WEVRF = \SI[group-separator={,},group-digits=integer]{62636857.86}{\square\meter\per\square\second}$ based on the European EUVN\_DA GNSS/levelling data set from~\cite{Kenyeres2010} is utilized in equation~(\ref{eq:pot2}), giving $\Clev$. For the GNSS/geoid approach according to equations~(\ref{eq:gravP}) or~(\ref{eq:pot1}), the disturbing potential $T$ or the corresponding height anomaly values $\zeta$ are taken from the EGG2015 model, and the normal potential $U_0 = \SI[group-separator={,},group-digits=integer]{62636860.850}{\square\meter\per\square\second}$, associated with the surface of the underlying GRS80 (Geodetic Reference System 1980; see~\cite{Moritz2000}) level ellipsoid, is used, resulting in $\CGNSS$. Furthermore, the mean normal gravity values $\bar\gamma$ are also based on the GRS80 level ellipsoid; for further details, see~\cite{Denker2017}. Taking all this into account, leads to the following discrepancies between the geopotential numbers from the GNSS/geoid and the geometric levelling approach, defined in the sense $\Delta C = \CGNSS - \Clev$:
\vspace{0.5em}\\
%
\begin{tabular}[h]{ll}
	$\bullet$~PTB:		& \SI{-0.017}{gpu},\\
	$\bullet$~LNE-SYRTE:	& \SI{-0.109}{gpu},\\
	$\bullet$~NPL:		& \SI{-0.275}{gpu} (with ODN offset based on official EVRF2007 results),\\
			& \SI{-0.144}{gpu} (with ODN offset based on own estimate, see above),\\
	$\bullet$~INRIM:		& \SI{+0.019}{gpu},\\
	$\bullet$~LSM:		& \SI{-0.087}{gpu}.\\
\end{tabular}
\vspace{0.5em}\\
The above results show first of all that the two approaches differ at the few decimetre level over Europe, that the consideration of the new French and channel tunnel levelling leads to a better agreement, and that the implementation of the national height system offsets was done correctly, recalling, e.g., that the difference between the French and German zero level surfaces is about half a metre. However, as the above differences $\Delta C$ are directly depending on the chosen reference potential $\WEVRF$ for EVRF2007, potential differences between two stations and the corresponding discrepancies between the GNSS/geoid and the geometric levelling approach are discussed as well in the following. Regarding potential differences, the discrepancies between both approaches amount to \SI{-0.106}{gpu} for the connection INRIM/LSM, \SI{-0.036}{gpu} for INRIM/PTB, \SI{-0.092}{gpu} for PTB/LNE-SYRTE, \SI{-0.166}{gpu} for LNE-SYRTE/NPL (\SI{-0.035}{gpu} based on own ODN offset, see above), as well as \SI{-0.258}{gpu} for PTB/NPL (\SI{-0.127}{gpu} based on own ODN offset, see above), respectively.

Regarding the significance of the aforementioned discrepancies in the potential differences between both geodetic approaches (levelling, GNSS/geoid), these have to be discussed in relation to the corresponding uncertainties of levelling, GNSS, and the quasigeoid model. Denker et al.~\cite{Denker2017} discuss the uncertainties (standard deviation) from single line levelling connections and the EVRF2007 network adjustment, indicating a factor 2.5 improvement due to the network adjustment. The EVRF2007 network adjustment gives a standard deviation of about \SI{20}{\mm} for the height connection PTB/LNE-SYRTE, while the corresponding standard deviations for the connections PTB/NPL and LNE-SYRTE/NPL are both about \SI{80}{\mm} (M. Sacher, BKG, Leipzig, Germany, personal communication, 10 May 2017), the latter being dominated by the uncertainty of the hydrodynamic levelling across the English Channel. However, these internal uncertainty estimates from the network adjustment do not consider any systematic levelling error contributions. On the other hand, the GNSS ellipsoidal heights have uncertainties below \SI{10}{\mm}, the uncertainty of EGG2015 has been discussed above, yielding a standard deviation of \SI{19}{\mm} for the absolute values and about \SI{27}{\mm} for corresponding differences over longer distances, and therefore some of the larger discrepancies between the two geodetic approaches (levelling versus GNSS/geoid) have to be considered as statistically significant. Hence, as systematic errors in levelling at the decimetre level exist over larger distances in the order of \SI{1000}{\km} (e.g.~in France, UK, and North America; see above), it is hypothesized that the largest uncertainty contribution to the discrepancies between both geodetic approaches comes from geometric levelling (see also~\cite{Denker2017}). Consequently, geometric levelling is recommended mainly for shorter distances of up to several ten kilometres, where it can give millimetre uncertainties, while over long distances, the GNSS/geoid approach should be a better approach than geometric levelling, and it can also deliver absolute potential values needed for contributions to international timescales.

\subsection{Unified relativistic redshift corrections}

\begin{figure}[!t]
\includegraphics[width=\linewidth]{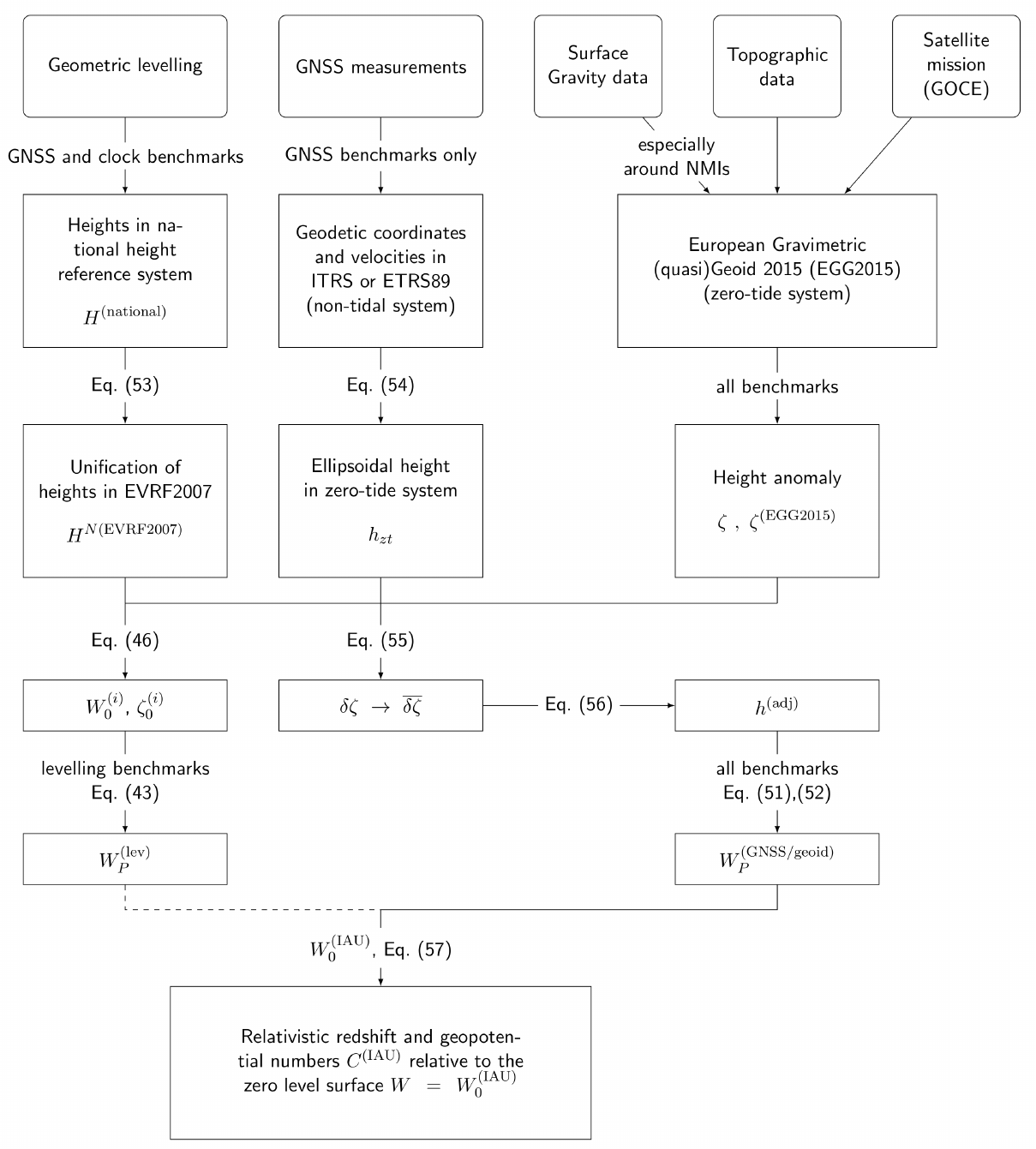}
\caption{Flowchart: from measurements to the determination of a unified relativistic redshift clock correction at European scale. It is possible to extend this chart to the worldwide scale wherever a high quality gravimetric model of the geoid exists.}
\label{fig:flowchart}
\end{figure}

The results from the gravity potential determination from both the geometric levelling and the GNSS/geoid approach are given in Table~\ref{tab:corr} for the two ITOC sites PTB and LNE-SYRTE as typical examples; further results for the other ITOC sites are foreseen for a separate publication, and corresponding results for further sites in Germany are documented in~\cite{Denker2017}. Based on the discussion in the preceding section as well as section~\ref{sec:unc}, Table~\ref{tab:corr} gives the relativistic redshift corrections only for the GNSS/geoid approach, which can be considered as the recommended values. The redshift corrections are based on the conventional value $\WIAU$, following the IERS2010 conventions and the IAU resolutions for the definition of TT (see Sect.~\ref{sec:TT}), using the equation~(\ref{eq:corr2}).
The uncertainty of the given relativistic redshift corrections based on the GNSS/geoid approach amounts to about $2\times10^{-18}$ (see above). All operations from the measurements to the final values of the unified relativistic redshift corrections are summarized in the flowchart given in figure~\ref{fig:flowchart}.

\begin{table}[t]
\centering
\caption{Ellipsoidal coordinates (latitude, longitude, height; $\phi$, $\lambda$, $h^{(adj.)}$) referring to ITRF2008 reference frame (epoch 2005.0; GRS80 ellipsoid; zero-tide system), normal heights $\HNevrf$ based on EVRF2007, geopotential numbers based on the geometric levelling ($\Clev$) and GNSS/geoid approach ($\CGNSS$) relative to the IAU2000 conventional reference potential $\WIAU$ and differences $\Delta C$ thereof, as well as the relativistic redshift correction based on the GNSS/geoid approach}
\label{tab:corr}
\scriptsize
\begin{tabular}{|l||l|l|l|l|l|l|l|l|}
\hline
Station & \multicolumn{3}{l|}{$\phi$} & \multicolumn{3}{l|}{$\lambda$} & $h^{(adj.)}$ & $\HNi$ \\
 & [\degre] & ['] & ["]  & [\degre] & ['] & ["]  & [m] & [m] \\ \hline
\multicolumn{9}{c}{\bf PTB, Braunschweig, Germany}\\ \hline
	PTBB & 52      & 17      & 46.28177 & 10            & 27      & 35.08676      & 130.201        & 87.442           \\ \hline
	LB03       & 52      & 17      & 49.94834 & 10            & 27      & 37.63590      & 143.514        & 100.758 \\ \hline
	AF02       & 52      & 17      & 30.90851 & 10            & 27      & 28.21874      & 123.716        & 80.945  \\ \hline
	MB02       & 52      & 17      & 47.22270 & 10            & 27      & 50.49262      & 144.932        & 102.173 \\ \hline
	KB01       & 52      & 17      & 45.2     & 10            & 27      & 33.1          & 119.627        & 76.867  \\ \hline
	KB02       & 52      & 17      & 46.3     & 10            & 27      & 35.1          & 119.708        & 76.949  \\ \hline
\multicolumn{9}{c}{\bf LNE-SYRTE, Paris, France}\\ \hline
	100        & 48      & 50      & 7.99682  & 2             & 20      & 8.38896       & 105.652        & 61.394  \\ \hline
	A          & 48      & 50      & 10.90277 & 2             & 20      & 10.55555      & 130.964        & 86.706  \\ \hline
	OPMT & 48      & 50      & 9.31198  & 2             & 20      & 5.77891       & 122.546        & 78.288  \\ \hline
\end{tabular}

\vspace{1em}

\begin{tabular}{|l||l|l|l||l|}
\hline
Station & $\Clev$ & $\CGNSS$ & $\Delta C$ & Redshift \\
 & [10~m$^2$s$^{-2}$] & [10~m$^2$s$^{-2}$] & [10~m$^2$s$^{-2}$] & [$10^{-16}$]\\ \hline
\multicolumn{5}{c}{\bf PTB, Braunschweig, Germany}\\ \hline
	PTBB & 85.617         & 85.600      & -0.017 & -95.243  \\ \hline
	LB03       & 98.684         & 98.667      & -0.017 & -109.782 \\ \hline
	AF02       & 79.242         & 79.225      & -0.017 & -88.150  \\ \hline
	MB02       & 100.072        & 100.055     & -0.017 & -111.326 \\ \hline
	KB01       & 75.241         & 75.224      & -0.017 & -83.698  \\ \hline
	KB02       & 75.321         & 75.304      & -0.017 & -83.787  \\ \hline
\multicolumn{5}{c}{\bf LNE-SYRTE, Paris, France}\\ \hline
	100        & 60.039         & 59.930      & -0.109 & -76.845  \\ \hline
	A          & 84.868         & 84.759      & -0.109 & -76.851  \\ \hline
	OPMT       & 76.611         & 76.502      & -0.109 & -81.712  \\ \hline
\end{tabular}
\end{table}

Finally, as the results from the geometric levelling approach and the GNSS/geoid approach are presently inconsistent at the decimetre level across Europe, the more or less direct observation of gravity potential differences through optical clock comparisons (with targeted fractional accuracies of~$10^{-18}$, corresponding to \SI{1}{\cm} in height) is eagerly awaited as a means for resolving the existing discrepancies between different geodetic techniques and remedying the geodetic height determination problem over large distances. A first attempt in this direction was the comparison of two strontium optical clocks between PTB and LNE-SYRTE via a fibre link, showing an uncertainty and agreement with the geodetic results of about $5\times10^{-17}$~\cite{Lisdat2016}. This was mainly limited by the uncertainty and instability of the participating clocks, which is likely to improve in the near future. 

Furthermore, for clocks with performance at the $10^{-17}$ level and below, time-variable effects in the gravity potential, especially solid Earth and ocean tides, have to be considered and can also serve as a method of evaluating the performance of the optical clocks (i.e.~a detectability test). Recent analysis of optical clock comparisons already included temporal variations~\cite{Takano2016,Delva2017e}. Then, after further improvements in the optical clock performance, conclusive geodetic results can be anticipated in the future, and clock networks may also contribute to the establishment of the International Height Reference System (IHRS).

\section{Contribution of chronometric geodesy to the determination of the geoid}

\label{sec:high_res}


In geodesy, geoid determination is understood as the determination of the shape and size of the geoid with respect to a well-defined coordinate reference system, which usually means the determination of the height of the geoid (geoid height) above a given reference ellipsoid. The problem is solved within the framework of potential theory and GBVPs, where the task is to find a harmonic function (i.e. the disturbing potential $T$) everywhere outside the Earth's masses (possibly after mass displacements and reductions), which fulfills certain boundary conditions. In principle, all measurements that can be mathematically linked to the disturbing potential $T$ (e.g. gravity anomalies, vertical deflections, gradiometer observations, and point-wise disturbing potential values itself), can contribute to the solution, but in practice gravity measurements usually play the main role in combination with topographic and global satellite gravity information (also denoted as the gravimetric method, see above). A very flexible approach, with the capability to combine all the aforementioned (inhomogeneous) measurements of different kinds and the option to predict (output) heterogeneous quantities related to $T$, is the least-squares collocation (LSC) method~\cite{Moritz1980}.

Regarding the use of clocks for gravity field modelling and geoid determination, this always implies that also precise positions of the clock points with respect to a well-defined reference system are required. This concerns mainly the ellipsoidal height, which should be available with the same (or lower) uncertainty than the clock-based physical heights or potential values, such that gravity field related quantities $N = h - H$ or $\zeta = h - H_N$ (cf. equation~(\ref{eq:undulation})) can be obtained, establishing a direct link to the disturbing potential $T$ (e.g. through equation~(\ref{eq:anomaly})); this is exactly the same situation as a combination of GNSS and geometric levelling (so-called GNSS/levelling), as employed since many years (e.g.~\cite{Denker1988,Denker1998}). Consequently, always clock plus GNSS measurements are required for gravity field modelling and geoid determination. Furthermore, in view of further improved clocks at (or below) the $10^{-18}$ level, it should be noted that an ellipsoidal height uncertainty of 5 to 10~mm is about the limit of what is achievable with GNSS today, requiring static and sufficiently long observation sessions and an appropriate post-processing. Clock measurements alone are directly equivalent to the results from geometric levelling and gravity measurements and hence can be considered as a height (but not a geoid) determination technique; if clocks can be compared with a (space) reference clock with known potential value, then this could help to realize the geoid, i.e. to find its position with respect to a given measurement point on the Earth's surface, but this still does not mean that one would know the coordinates of the corresponding geoid point (i.e. its ellipsoidal height or geoid height).


Distant clock comparisons and GNSS measurements provide a new kind of geodetic observable, which is complementary to the classical geodetic measurements (terrestrial and satellite gravity field observations). We have seen in section~\ref{sec:GNSS_geoid} that satellite and terrestrial data (mainly gravity and topography) complement each other, with the satellite data providing the long wavelength field structures, while the terrestrial data contributes to the short wavelength features. Indeed, terrestrial data (gravity and topography) is most sensitive to small-scale spatial variations of the gravity potential. For this reason, insufficiently dense terrestrial data can lead to significant errors in the determination of the geoid. 

By nature, potential data are smoother and more sensitive to mass sources at large scales than gravity data. They can complement the information given by the gravity data in the same way as the satellite data does, but on smaller scales. Therefore they could provide the medium wavelength field structure, in between the spectral information of classical terrestrial data and satellite data. They could reduce the error in the determination of the geoid where gravity data is too sparse to reconstruct the medium wavelengths field structures. Indeed, gravity data are sometimes sparsely distributed: the plains are generally densely surveyed, while the mountainous regions are poorly covered because several areas are mostly inaccessible by conventional gravity surveys. Clock and GNSS data nearby these inaccessible areas could reduce the error in the determination of the geoid.


To illustrate the potential benefits of clocks and GNSS in geodesy, the determination of the geopotential at high spatial resolution, about 10~km, was investigated in~\cite{Lion2017}. The tested region is the Massif Central in France. It is interesting because it is characterized by smooth, moderate altitude mountains and volcanic plateaus, leading to variations of the gravitational field over a range of spatial scales. In such type of region, the scarcity of gravity data is an important limitation in deriving accurate high resolution geopotential models. 


\subsection{Methodology}%
\begin{figure}[t]
	\centering
	\includegraphics[width=0.8\linewidth]{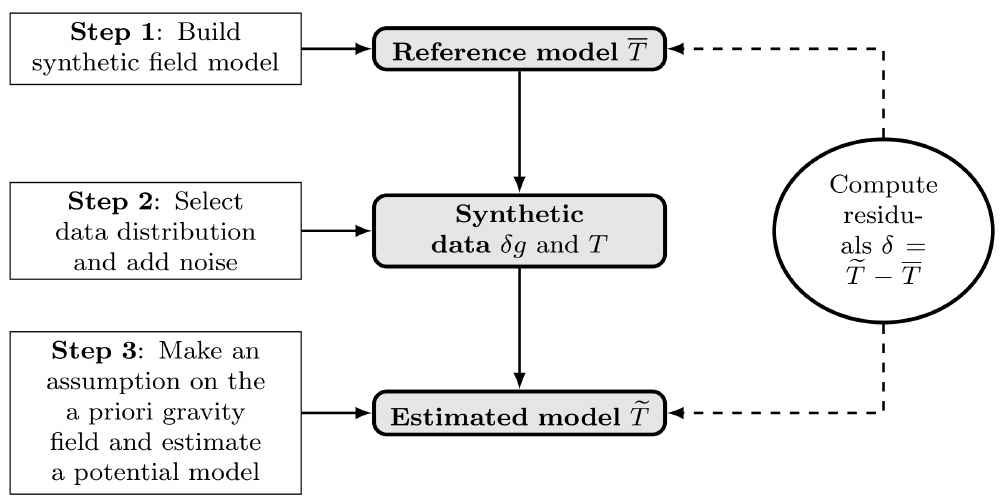}
	\caption{Scheme of the numerical approach used to evaluate the contribution of atomic clocks.}
	\label{fig:methodo}
\end{figure}

The simulations are based on synthetic data (gravity and potential/clock data) and consist in comparing the quality of the geopotential reconstruction solutions from the gravity data, with or without taking into account clock data. In the following, ``clock data'' is considered as disturbing potential $T$ values derived from clock and GNSS measurements as outlined above (on the basis of equation~(\ref{eq:anomaly})). The synthetic gravity and potential data are sampled by using a state of the art geopotential model \cite[EIGEN-6C4]{Forste2014} up to degree and order 2000 (i.e.~10~km resolution), and some spatial distribution of points. The solutions are estimated thanks to an inversion method, requiring a covariance model to interpolate the data, and they are compared to a reference model. In more details, the numerical process is presented below and sketched up in Figure~\ref{fig:methodo}:
\begin{enumerate}
	\item Step 1: Generation of the reference model of the disturbing potential~$\Xref{T}$ with program GEOPOT~\cite{Smith1998}, which allows to compute the gravity field related quantities at given locations by using mainly a geopotential model. The long wavelengths of the gravity field covered by the satellites and longer than the extent of the local area are removed, providing centered or close to zero data for the determination of a local covariance function. The terrain effects are removed with the help of the topographic potential model dV\_ELL\_RET2012 \cite{Claessens2013};
	\item Step 2: Generation of the synthetic data~$\grdist$ and~$T$ from a realistic spatial distribution. A white noise is then added to $\grdist$ and~$T$, with a standard deviation of \pot{0.1} (i.e.~\SI{1}{\centi\meter} on the geoid) for clocks and 1~mGal for gravimetric measurements;
	\item Step 3: Estimation of the disturbing potential~$\Xest{T}$ from the synthetic data~$\grdist$ only and then in combination with the synthetic data $T$ on the 10-km grid using the Least-Squares Collocation (LSC) method. In this step, a logarithmic 3D covariance function is employed~\cite{Forsberg1987}. This 
	model has the advantage to provide the auto-covariances (ACF) and cross-covariances (CCF) of the potential~$T$ and its derivatives in closed-form expressions. Parameters of this model are adjusted to the empirical ACF of~$\grdist$ with the program GPFIT \cite{Forsberg2003}. Note that low frequencies are included in this covariance function, not removed as done in step 1.
	\item Final step : Evaluation of the potential recovery quality for selected data situations by comparing the statistics of the residuals~$\delta$ between the estimated values~$\Xest{T}$ and the reference model~$\Xref{T}$.
\end{enumerate}

Let us underline that in this work, we use synthetic potential data while a network of clocks would give access to potential differences between the clocks. We indeed assume that the clock-based potential differences have been connected to one or a few reference points, without introducing additional biases larger than the assumed clock uncertainties. In order to have more realistic simulations, we should add the noise due to uncertainty on the geometric coordinates of the clock, especially the vertical component. This is a work in progrees. However, if this error is below the accuracy of the clock, i.e.~\pot{0.1} (\SI{1}{\centi\meter} on the geoid), they will not change the main conclusions of this work.

\subsection{Data set}%
\begin{figure}[t]
	\centering
	\includegraphics[clip=true, trim = 0 0 0 0, height=0.25\paperheight]{%
		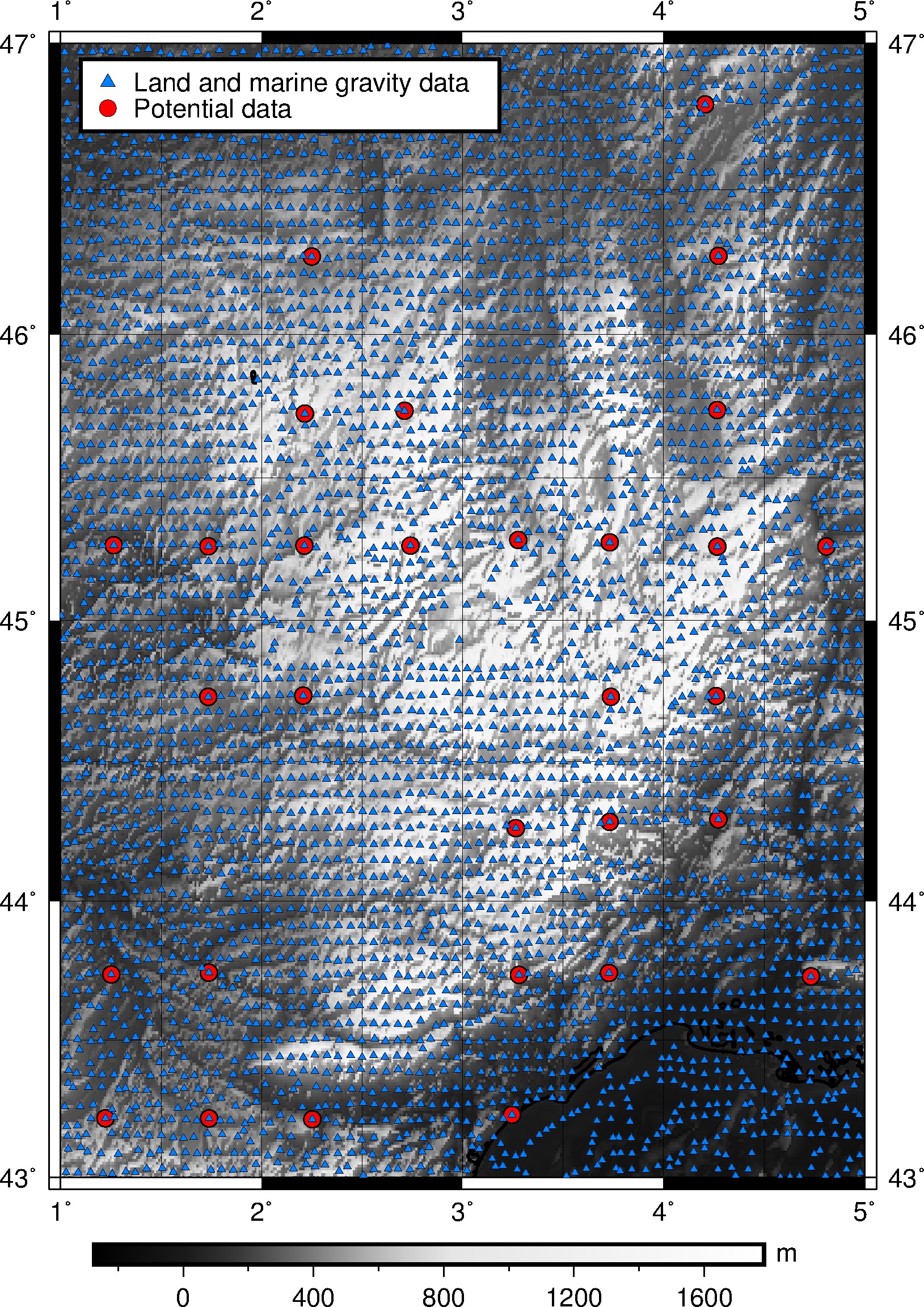}
	
	\caption{Spatial distribution of \num{4374}~gravity data and \num{33}~clock data used in the synthetic tests.}
	\label{fig:Coverage_Tdg}
\end{figure} 
The locations of the gravimetric data are chosen to reproduce a realistic distribution of measurements. Their spatial distribution can be obtained from the BGI (International Gravimetric Bureau) database, then under-sampled by using a data reduction process, as plotted in blue in Figure \ref{fig:Coverage_Tdg}.
For this test case, the clock measurements (red markers) are put only where existing land gravity data are located and in areas where the gravity data coverage is poor. Moreover, in order to avoid clock points to be too close to each other, a minimal distance is defined between them.

\subsection{Contribution of clocks}%
\begin{figure*}[t]
	\centering
	\null\vfill
	\subfloat[Without clock data.]{%
		\raisebox{0\height}{\includegraphics[clip=true, trim = 30 0 30 0, width=0.9\linewidth]{%
				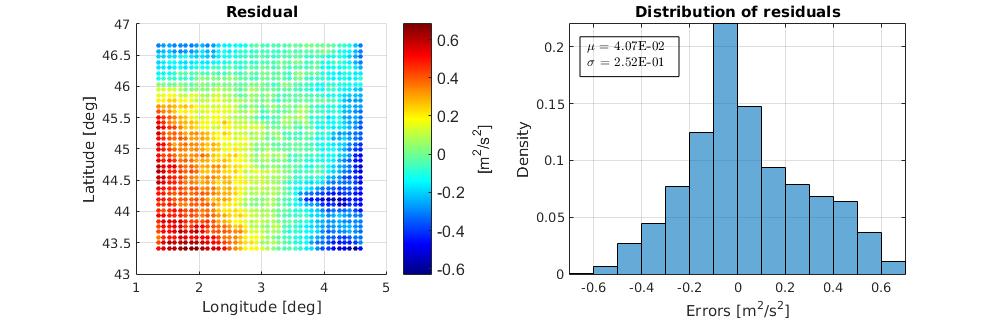}}
		\label{fig:Auv_EMC_Tdg_from_T0dg_4art}}
	\vfill
	\subfloat[With clock data.]{%
		\includegraphics[clip=true, trim = 30 0 30 0, width=0.9\linewidth]{%
			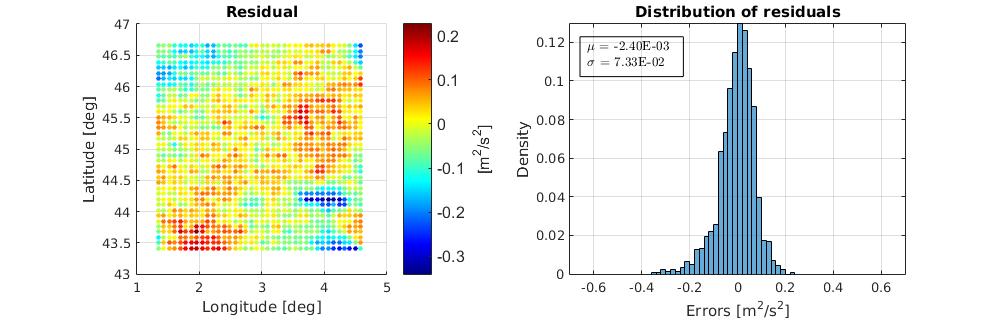}
		\label{fig:Auv_EMC_Tdg_from_T33dg_56_15_4art}}
	\vfill\null
	
	\caption{Accuracy of the disturbing potential~$T$ reconstruction on a regular 10-km grid in Massif Central, obtained by comparing the reference model and the reconstructed one. In Figure (a), the estimation is realized from the \num{4374} gravimetric data~$\delta g$ only, and in Figure (b) by adding 33 potential data~$T$ to the gravity data. To avoid edge effects in the estimated potential recovery, a grid edge cutoff of 30~km has been removed in the solutions.}
	\label{fig:Auv_T_recovery}
\end{figure*} 
In Figure \ref{fig:Auv_T_recovery}, it is shown that adding the clock-based potential values to the existing gravimetric data set can notably improve the reconstruction of the potential~$T$. In Figure~\ref{fig:Auv_EMC_Tdg_from_T0dg_4art}, the \num{4374}~gravimetric data are used as input and the disturbing potential is estimated with a bias~$\mu_T \approx \pot{0.041}$ (4.1~mm) and a rms~$\sigma_T \approx \pot{0.25}$ (2.5~cm). By combining the gravimetric measurements and the 33~potential measurements, see  Figure~\ref{fig:Auv_EMC_Tdg_from_T33dg_56_15_4art}, the bias is improved by one order of magnitude ($\mu_T \approx \pot{-0.002}$ or~\SI{-0.2}{\mm}) and the standard deviation by a factor~3 ($\sigma_T \approx \pot{0.07}$ or 7~mm). From the comparison of Figures~\ref{fig:Auv_EMC_Tdg_from_T0dg_4art} and~\ref{fig:Auv_EMC_Tdg_from_T33dg_56_15_4art} it is clear that the pure gravimetric solution exhibits a significant trend, which may be related to the data collection area and covariance function used, while the additional potential data effectively remove this trend.

Another important conclusion stemming from our simulations is that in solving the problem of gravity field recovery it is not required to have a dense clock network. As shown in~\cite{Lion2017} only a very few percent of clock measurements compared to the number of needed gravity data is sufficient. A more detailed study discussing the role of different parameters, such as the noise level in the data, effects of the resolution of gravity measurements and modeling errors can be found in \cite{Lion2017}. 

As a result of this work, ways to optimize clock location points has begun in order to answer to a practical question: where to put the geopotential measurements to minimize the residuals and improve further the determination of the gravity field? This is important when the gravimetric measurements can be tarnished by correlated errors. For this, we have implemented the optimization of a spatial distribution of clocks completing a pre-existing gravimetric network, by using the genetic algorithm $\epsilon$-MOEA (Multi Objective Evolutionary Algorithm, see \cite{Coulot2015}). 

\section{Conclusions}

We presented in this chapter what is chronometric geodesy, introducing notions and methods, both theoretical and experimental. The interest in this rather new topic is raised by the tremendous ameliorations of atomic clocks in the last decade; it is at the crossroads of general relativity and physical geodesy. On the one hand, physical geodesy is essential in order to model the relativistic redshift in distant clock comparisons, as well as to establish global timescales such as the TAI. On the other hand, when the limitations of physical geodesy are reached in term of method inaccuracies, then the clock comparison observable, which gives directly gravity potential differences, could bring something new for physical geodesy.

The question whether these ideas will emerge one day as operational methods depends on a large part on technological challenges. The development of sufficiently accurate and transportable optical clock is not a barrier, and several projects go in this direction~\cite{Koller2017,Cao2017,Yasuda2017}. The frequency transfer method is more challenging, especially on global scales (see section~\ref{sec:tech}). Optical fibres transfer fully meets the expectations of current and future optical clocks, but are limited to continental scales and are available only along predefined paths. Phase coherent free space optical links are being developed, but are currently limited by the effect of atmospheric turbulence. This method would be more adapted to global scales, especially if we think about some islands in the middle of the ocean, which are unlikely to be linked with an optical fibre.

Finally, we have to speak about the stability and integration <F2>time of optical clocks. Some recent techniques such as three-dimensional optical lattice clocks~\cite{Campbell2017} allow to greatly improve the integration times necessary to attain some specified accuracy for the clock. This would permit, in a distant comparison, to obtain the variations of the gravity potential with a good time resolution, and could lead to new ideas for the study of geophysical phenomenon.

\section*{Acknowledgments}
The authors would like to thank J\'er\^ome Lodewyck (SYRTE/Paris Observatory) for providing Figure~\ref{fig:accuracy}, and Martina Sacher (Bundesamt f\"ur Kartographie und Geod\"asie, BKG, Leipzig, Germany) for providing information on the EVRF2007 heights and uncertainties, the associated height transformations, and a new UELN adjustment in progress.

This research was supported by the European Metrology Research Programme (EMRP) within the Joint Research Project ``International Timescales with Optical Clocks'' (SIB55 ITOC), as well as the Deutsche Forschungsgemeinschaft (DFG) within the Collaborative Research Centre 1128 ``Relativistic Geodesy and Gravimetry with Quantum Sensors (geo-Q)'', project C04. The EMRP is jointly funded by the EMRP participating countries within EURAMET and the European Union. We gratefully acknowledge financial support from Labex FIRST-TF and ERC AdOC (Grant No. 617553).

\bibliographystyle{unsrt}
\bibliography{delva_relgeo_proceedings_2016}

\end{document}